\documentclass[twocolumn]{aastex62}

\newcommand{\Msun}{$M_{\odot}$}

\newcommand{\Av}{A$_V$}

\newcommand{\mic}{$\mu$m}

\newcommand{\Mearth}{$M_\oplus$}
\usepackage{placeins}
\shorttitle{ALMA Observations of L1641}
\shortauthors{Grant et al.}

\begin{document}

\title{An ALMA Survey of Protoplanetary Disks in Lynds 1641}

\correspondingauthor{Sierra Grant}
\email{sierrag@bu.edu}

\author[0000-0002-4022-4899]{Sierra L. Grant}
\affil{Institute for Astrophysical Research, Department of Astronomy, Boston University, 725 Commonwealth Ave., Boston, MA, USA}

\author[0000-0001-9227-5949]{Catherine C. Espaillat}
\affil{Institute for Astrophysical Research, Department of Astronomy, Boston University, 725 Commonwealth Ave., Boston, MA, USA}

\author[0000-0002-6808-4066]{John Wendeborn}
\affil{Institute for Astrophysical Research, Department of Astronomy, Boston University, 725 Commonwealth Ave., Boston, MA, USA}

\author[0000-0002-6195-0152]{John J. Tobin}
\affil{National Radio Astronomy Observatory, 520 Edgemont Rd., Charlottesville, VA 22903, USA}

\author[0000-0003-1283-6262]{Enrique Mac\'{i}as}
\affil{Joint ALMA Observatory, Alonso de Córdova 3107, Vitacura, Santiago 763-0355, Chile}
\affil{European Southern Observatory (ESO), Alonso de Córdova 3107, Vitacura, Santiago 763-0355, Chile }

\author[0000-0002-3091-8061]{Anneliese Rilinger}
\affil{Institute for Astrophysical Research, Department of Astronomy, Boston University, 725 Commonwealth Ave., Boston, MA, USA}

\author[0000-0003-3133-3580]{\'{A}lvaro Ribas}
\affil{European Southern Observatory (ESO), Alonso de Córdova 3107, Vitacura, Santiago 763-0355, Chile }

\author[0000-0001-7629-3573]{S. Thomas Megeath}
\affil{Ritter Astrophysical Research Center, Department of Physics and Astronomy, University of Toledo, Toledo, OH 43606, USA}

\author[0000-0002-3747-2496]{William J. Fischer}
\affil{Space Telescope Science Institute, Baltimore, MD 21218, USA}

\author[0000-0002-3950-5386]{Nuria Calvet}
\affil{Department of Astronomy, University of Michigan, Ann Arbor, MI 48109, USA}

\author[0000-0001-9597-7196]{Kyoung Hee Kim}
\affil{Korea Astronomy and Space Science Institute, Korea Astronomy and Space Science Institute (KASI), 776, Daedeokdae-ro, Yuseong-gu, Daejeon 305-348, Republic of Korea}

\begin{abstract}
We present ALMA observations of 101 protoplanetary disks within the star-forming region Lynds 1641 in the Orion Molecular Cloud A. Our observations include 1.33 mm continuum emission and spectral windows covering the J=2-1 transition of $^{12}$CO, $^{13}$CO, and C$^{18}$O. We detect 89 protoplanetary disks in the dust continuum at the 4$\sigma$ level ($\sim$88\% detection rate) and 31 in $^{12}$CO, 13 in $^{13}$CO, and 4 in C$^{18}$O. Our sample contains 23 transitional disks, 20 of which are detected in the continuum. We target infrared-bright Class II objects, which biases our sample towards massive disks. We determine dust masses or upper limits for all sources in our sample and compare our sample to protostars in this region. We find a decrease in dust mass with evolutionary state. We also compare this sample to other regions surveyed in the (sub-)millimeter and find that Lynds 1641 has a relatively massive dust disk population compared to regions of similar and older ages, with a median dust mass of 11.1$^{+32.9}_{-4.6}$ \Mearth\ and 27\% with dust masses equal to or greater than the minimum solar nebula dust mass value of $\sim$30 \Mearth. We analyze the disk mass-accretion rate relationship in this sample and find that the viscous disk lifetimes are similar to the age of the region, however with a large spread. One object, [MGM2012] 512, shows large-scale ($>$5000 AU) structure in both the dust continuum and the three gas lines. We discuss potential origins for this emission, including an accretion streamer with large dust grains.

\end{abstract}

\section{Introduction}
Protoplanetary disks contain the reservoirs of dust and gas that go into assembling planets. To build planets, a disk must contain enough mass to form planetary cores, and the planet formation process must take place before the disk is depleted of material. Disk evolution proceeds relatively rapidly; by $\sim$10 Myr, disks are found around only 20\% of low-mass stars and no high-mass stars \citep{hernandez07,ribas15}. Dust is a critical component to this evolution and it is a necessary ingredient in forming planetesimals, which are precursors to terrestrial planets and the cores of giant planets. The evolution of dust thus has a large impact on planet formation.

Millimeter and sub-millimeter wavelengths trace both dust continuum emission and important chemical signatures. Surveys of protoplanetary disks in the (sub-) millimeter have provided great insight into protoplanetary disk properties and evolution. These surveys include NGC 2024 (0.5 Myr, 414 pc, \citealt{vanTerwisga20}), ONC (1 Myr, 400 pc, \citealt{eisner18}), OMC1 (1 Myr, 400 pc, \citealt{eisner16}), OMC2 (1 Myr, 414 pc, \citealt{vanTerwisga19}), Lupus (1-3 Myr, 150-200 pc, \citealt{ansdell16}), Corona Australis (1-3 Myr, 160 pc, \citealt{cazzoletti19}), Ophiuchus (1 Myr, 140 pc, \citealt{cieza19, williams19}), Chamaeleon I (2 Myr, 160 pc, \citealt{pascucci16}), Taurus (2 Myr, 140 pc, \citealt{andrews13}), IC 348 (2-3 Myr, 310 pc, \citealt{ruizrodriguez18}), $\sigma$ Orionis (3-5 Myr, 385 pc, \citealt{ansdell17}), $\lambda$ Orionis (5 Myr, 400 pc, \citealt{ansdell20}), and Upper Scorpius (5-11 Myr, 145 pc, \citealt{barenfeld16}). The results of these works have shown that there is a dependence of disk mass on stellar mass (e.g., \citealt{andrews13,pascucci16,ansdell17,eisner18}), the age of the region (e.g., \citealt{barenfeld16}), and the evolutionary state \citep{tobin20}, and that the star-forming environment has an impact on disk evolution \citep{eisner18}. However, some star-forming regions show dust masses that are inconsistent with a simple monotonic decrease with age. The Ophiuchus \citep{williams19} and Corona Australis \citep{cazzoletti19} regions have lower disk masses than slightly older counterparts, indicating that our understanding of disk evolution is not yet complete.

To tease out the effects of environment and protoplanetary disk phase on disk evolution, it is beneficial to look at a single star-forming region with large populations of disks at a range of evolutionary stages. Lynds 1641 (L1641) provides such a laboratory. L1641 is a $\sim$1.5 Myr star-forming region within the Orion A molecular cloud (with declinations from -6$^{\circ}$ to -9$^{\circ}$), the nearest giant molecular cloud complex. \cite{kounkel17} used the Very Long Baseline Array to determine that the Southern end of L1641 lies at a distance of 428 $\pm$ 10 pc, although the filament has a distance gradient such that the northern end lies closer. This was also found using Gaia DR2 data by \citet{grossschedl18}. The \cite{megeath12,megeath16} \textit{Spitzer} survey of Orion identified over 700 young stars with disks in L1641. L1641 also contains 165 protostars that were recently observed as part of the Orion VLA/ALMA Nascent Disk and Multiplicity (VANDAM) survey \citep{tobin20}. Additionally, L1641's distance from the Orion Nebula Cluster means that it lacks nearby massive irradiating OB stars, making it comparable to surveys of low-density regions like Taurus and Lupus. However, it can also complement other regions in Orion, including dense, irradiated environments like the ONC and OMC1, dense but low-irradiation environments like OMC2, and low-density, irradiated environments like $\sigma$ Ori. The stellar population in L1641 has also been well-studied, including the determination of spectral types, accretion rates, and the analysis of X-ray observations \citep{fang09,fang13,hsu12,pillitteri13}. L1641 offers a large population of disks spanning a range of evolutionary states in a single region, making it an ideal laboratory for testing disk evolution theories.

Here we present ALMA observations of 101 young disks observed in L1641. This large sample of Class II objects (i.e., a pre-main sequence star surrounded by a disk with a spectral index of $-1.6<n_{[4.5]-[24]}<-0.3$, where $n=d$log$(\lambda F_{\lambda})/d$log$(\lambda)$, e.g., \citealt{furlan16}) complements previous (sub-)millimeter surveys of star-forming regions due to its young age and proximity to a large number of objects in younger evolutionary states for comparison. The L1641 sample presented here was selected from the \cite{grant18} sample of far-infrared-bright targets observed by \textit{Herschel} that lie along the dense L1641 filament. L1641 has been well-characterized out to the far-infrared wavelengths and it already shows signs of dust evolution and has a large fraction (23\%) of transitional disks (disks with gaps or cavities, \citealt{grant18}). This work seeks to characterize these disks at radio wavelengths.

In Section~\ref{sec: sample} we discuss the sample and in Section~\ref{sec: observations} we present the ALMA observations. In Section~\ref{sec: results} we describe the continuum fluxes, gas fluxes, and compare the dust masses derived from the continuum fluxes to other surveys done in the (sub-)mm. Additionally, in Section~\ref{sec: results}, we present observations of one object that shows large-scale structure in the dust continuum and gas lines. We discuss our results in Section~\ref{sec: discussion}. A summary is provided in Section~\ref{sec: summary and conclusions}.

\section{Sample}\label{sec: sample}

\begin{figure}
    \centering
    \includegraphics[scale=0.55]{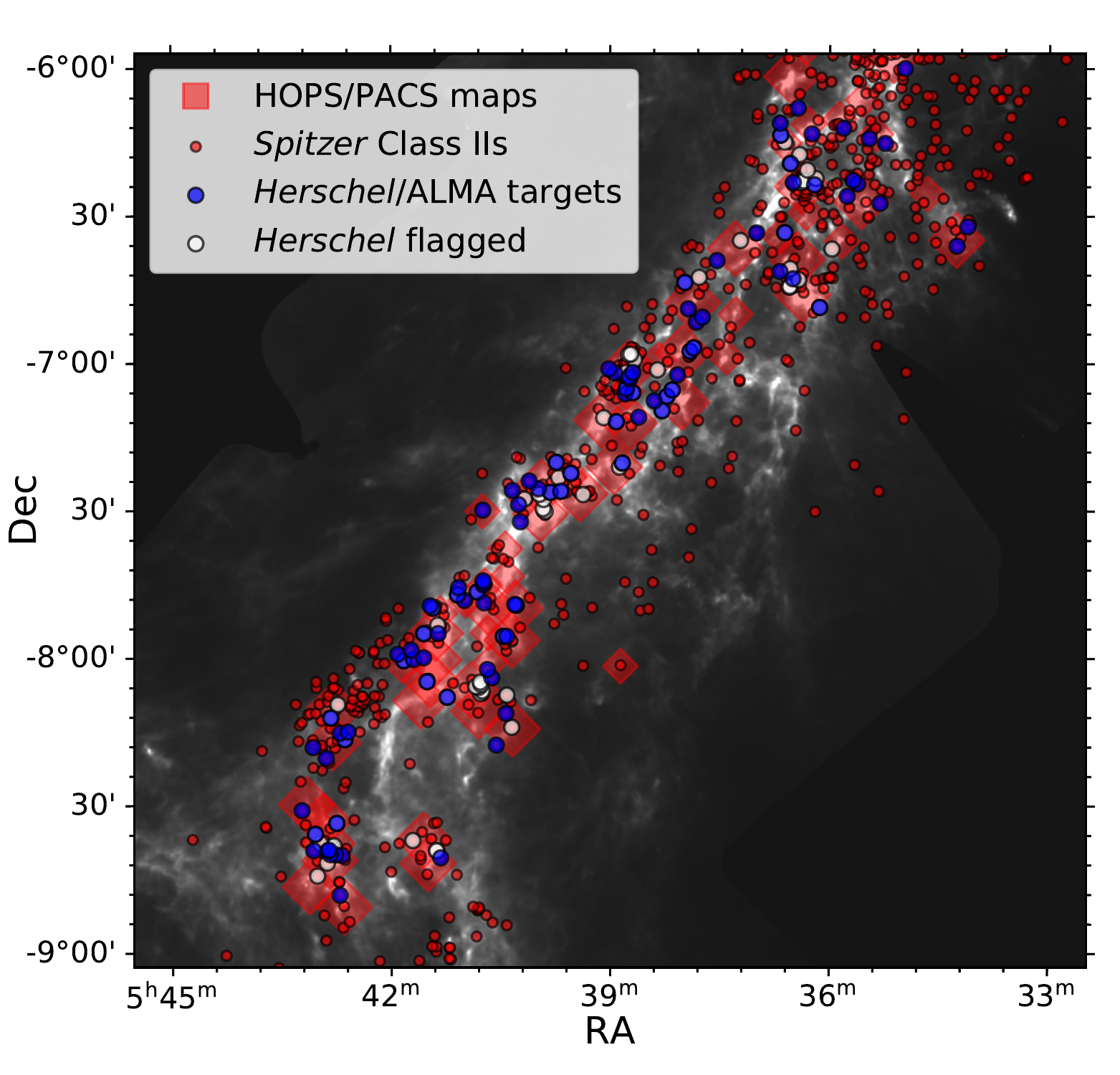}
    \caption{The Lynds 1641 target map. The \textit{Spitzer}-identified young stars with disks are shown as small red points \citep{megeath12}, the objects detected by \textit{Herschel} but flagged in \cite{grant18} are shown in white, and the \textit{Herschel} targets observed by ALMA are shown in blue. The approximate sizes (8'$\times$8' or 5'$\times$5') and locations of the HOPS/PACS maps are plotted in red squares, however, we note that the actual maps are not perfect squares. The background image is the L1641 column density, N(H), map shown on a log scale adapted from \cite{stutz&kainulainen15} and \cite{stutz&gould16}.}
    \label{fig: hops coverage}
    % from Limits.py
\end{figure}

\cite{megeath12} surveyed the entire Orion A Molecular Cloud with \textit{Spitzer} and identified 724 young stars with disks and 165 protostars in L1641 with declinations between -6.0$^\circ$ and -9.0$^\circ$. This region was observed as part of the \textit{Herschel} Orion Protostar Survey (HOPS; \citealt{fischer10, manoj13, stutz13, furlan16, fischer17, fischer20}). The HOPS survey targeted \textit{Spitzer}-identified protostars in Orion using the Photodetector Array Camera and Spectrometer (PACS; \citealt{poglitsch10}). The survey consisted of 5'$\times$5' or 8'$\times$8' maps targeting the protostars with expected 70 \mic\ flux densities greater than 42 mJy. These protostars lie largely in the dense filament of L1641. \citet{grant18} discussed the Class II sample that was serendipitously observed in the HOPS fields in L1641. We focus on the HOPS maps that covered L1641 (i.e., below -6$^\circ$; red squares Figure~\ref{fig: hops coverage}). Of the 724 Class IIs below -6$^\circ$ identified by \cite{megeath12}, 581 fall
into the HOPS maps and of those, only 180 targets were detected at 70 \mic\ (see white and blue points in Figure~\ref{fig: hops coverage}). Additionally, objects that had \textit{Herschel} fluxes that were flagged in \cite{grant18} are not included in the observations for this work. Objects were flagged due to 1) close sources or nebulosity in their \textit{Herschel} maps, 2) visual extinctions, \Av, equal to or greater than 15, or 3) colors uncharacteristic of classical T Tauri stars. 

The final \cite{grant18} sample contained 104 disks. Of those, only four are known binaries ([MGM2012] 250, 561, 579, and 980), although only $\sim$15\% of the sample had been surveyed for binarity. The majority of the existing binary coverage comes from the \textit{Hubble Space Telescope} survey by \cite{kounkel16}, which searched for companions between 100 and 1000 AU. For our ALMA observations, we remove binaries with either unknown separations or separations less than 800 AU ([MGM2012] 250, 579, and 980). Thus the ALMA sample of 101 disks contains no known binaries with small separations, although with the limited coverage, we cannot rule out that some objects are multiples.

As noted above, for our ALMA sample we only include those 101 disks that were not flagged (blue points in Figure~\ref{fig: hops coverage}), representing $\sim$14\% of the Class IIs in L1641 and $\sim$17\% of the Class IIs in L1641 that were covered by the HOPS maps. This leads to a bias in our sample selection, namely that the disks lie along the L1641 filament. \cite{megeath16} find that the protostar/pre-main-sequence star ratio decreases from clustered to distributed regions in L1641. Thus, it is likely that the clustered objects are younger than the distributed young stellar objects. This sample then likely contains younger objects than Class II sources that have dispersed outward from the filament and/or this sample lacks Class IIs that are formed in more isolated environments. Additionally, within the HOPS fields, not all Class IIs were detected by \textit{Herschel} and thus are not included in the ALMA sample we present here. With this in mind, the findings we present should be interpreted with caution. While the results may not be representative of L1641 as a whole, the sample is sizable and represents an interesting subset of objects: young, infrared-bright, and likely massive disks at the beginning of their Class II phases. 

\section{ALMA Observations}\label{sec: observations}
Our sample of 101 protoplanetary disks in L1641 was observed on 2019 December 12, 13, and 16 as part of Cycle 7 (Project ID: 2019.1.00951.S; PI: Grant). Calibration was done by the NAASC/ALMA staff with ALMA pipeline and CASA version 5.6.1-8. These observations were carried out in Band 6 with 43-44 antennas and baselines from 15 to 313 m. The time on each source was $\sim$24 seconds per observation date, adding up to a total of 72 seconds for each target for an average rms of 0.12 mJy beam$^{-1}$. Two continuum windows are included, each with a bandwidth of 1.875 GHz, centered at 231.986 and 216.987 GHz. Three spectral windows are included to observe the $^{12}$CO, $^{13}$CO, and C$^{18}$O J=2-1 transitions, each with a bandwidth of 0.117 GHz and centered at 230.524, 220.385, and 219.547 GHz, respectively. J0725-0054 was used as a bandpass and flux calibrator and J0542-0913 was used as a phase calibrator. All three observations were combined to improve signal-to-noise. 

Continuum images were cleaned in CASA version 5.6.1 using the \texttt{tclean} task using Briggs weighting with a robust parameter of +0.5 and the \texttt{mtmfs} deconvolver. The resulting images typically had beam sizes of $\sim$1.3''$\times$1.0'' and a background rms of 0.12 mJy beam$^{-1}$. The line data were first averaged by combining two adjacent channels to increase signal-to-noise. Then, the data were continuum-subtracted using the task \texttt{uvcontsub}. Next, the data were cleaned and imaged using the \texttt{tclean} task, using Briggs weighting and a robust parameter of +0.5. Zero-moment maps for our targets were constructed using the task \texttt{immoments}, clipping the data below 1$\sigma$, where 1$\sigma$ is equal to the rms of $\sim$20 emission-free channels on either end of the continuum-subtracted cube. The velocity range used for integration varied from target to target, depending on the range in which line emission was present. The rms was typically about 13, 13, and 11 Jy/beam for $^{12}$CO, $^{13}$CO, and C$^{18}$O, respectively. 

In addition to the planned targets, an additional 10 Class II objects from the \citet{megeath12} \textit{Spitzer} catalog fell into the ALMA maps we present here. These 10 objects have \citet{megeath12} ID numbers 234, 379, 412, 573, 591, 596, 731, 793, 918, and 995. Only one of these objects, [MGM2012] 918, was included in the \textit{Herschel} survey of \citet{grant18}, however, the object is flagged in that work for having nearby sources and/or bright nebulosity at 70 \mic, and therefore is not included in this ALMA sample. Of these 10 objects from \citet{megeath12}, four are detected in the continuum and these are listed in Table~\ref{tab: extras table} (\#2, 4, 5, and 8). An additional four objects are detected in the continuum, we list the coordinates of these additional objects in Table~\ref{tab: extras table} as well (\#1, 3, 6, and 7). Additional object \#6 (very near [MGM2012] 673) is also detected in the $^{12}$CO and $^{13}$CO. We do not include these additional objects in our analysis.

\FloatBarrier 
\startlongtable
\begin{deluxetable*}{cccccccccc}
\tablecaption{1.33 mm Continuum Properties \label{tab: observations table}}
% \tabletypesize{\scriptsize}
\tablehead{
\colhead{{[}MGM2012{]} ID} & \colhead{RA} & \colhead{Dec} & 
\colhead{SpT} & \colhead{Dist} & \colhead{F$_{1.33\ mm}$} & \colhead{M$_{dust}$} & \colhead{Major} & \colhead{Ratio} & \colhead{PA} \\
\colhead{} & \colhead{(J2000)} & \colhead{(J2000)} & \colhead{} & \colhead{(pc)} & \colhead{(mJy)} & \colhead{(M$_{\oplus}$)} & \colhead{(arcsec)} & \colhead{} & \colhead{($^\circ$)}
}
\colnumbers
\startdata
198&05:42:42.4&-08:48:13.8&M2&428$\pm$10$^a$&$<$0.48&$<$2.64&-&-&-   \\ 
217&05:41:19.6&-08:40:38.7&-&520$\pm$90&0.48$\pm$0.10&4$\pm$1&-&-&-   \\ 
223&05:42:40.8&-08:40:08.6&M2.5&420$\pm$20&4.35$\pm$0.10&23$\pm$2&-&-&-   \\ 
225&05:42:46.1&-08:40:00.8&K4&428$\pm$10$^a$&0.58$\pm$0.10&3.2$\pm$0.6&-&-&-   \\ 
227&05:42:50.5&-08:39:57.5&K0&400$\pm$40&2.95$\pm$0.10&14$\pm$3&-&-&-   \\ 
228&05:42:52.5&-08:39:16.6&-&420$\pm$20&5.50$\pm$0.10&29$\pm$3&-&-&-   \\ 
231&05:43:03.9&-08:39:09.2&M1.5&428$\pm$10$^a$&1.09$\pm$0.10&6.0$\pm$0.6&-&-&-   \\ 
232&05:42:50.0&-08:39:02.8&-&428$\pm$10$^a$&0.56$\pm$0.10&3.1$\pm$0.6&-&-&-   \\ 
233&05:42:51.4&-08:39:01.9&-&428$\pm$10$^a$&$<$0.48&$<$2.64&-&-&-   \\ 
256&05:43:02.6&-08:35:48.8&-&428$\pm$10$^a$&2.50$\pm$0.10&13.8$\pm$0.8&-&-&-   \\ 
263&05:42:45.0&-08:33:36.3&M2.5&390$\pm$40&$<$0.48&$<$2.15&-&-&-   \\ 
269&05:43:13.5&-08:31:00.5&M3&600$\pm$300&13.4$\pm$0.1&100$\pm$200&0.22$\pm$0.04&1.0$\pm$0.3&0$\pm$70   \\ 
278&05:42:53.6&-08:20:22.6&K6&420$\pm$20&$<$0.48&$<$2.49&-&-&-   \\ 
282&05:43:04.4&-08:18:10.8&M2.5&460$\pm$40&0.61$\pm$0.10&3.9$\pm$0.9&-&-&-   \\ 
284&05:40:33.7&-08:17:43.4&-&700$\pm$1000&5.37$\pm$0.10&100$\pm$200&-&-&-   \\ 
291&05:42:38.2&-08:16:35.4&-&600$\pm$200&$<$0.48&$<$4.38&-&-&-   \\ 
294&05:42:42.1&-08:15:15.1&M3&430$\pm$30&0.84$\pm$0.10&4.6$\pm$0.9&-&-&-   \\ 
296&05:42:35.6&-08:15:01.8&M2&400$\pm$20&3.40$\pm$0.10&16$\pm$2&-&-&-   \\ 
307&05:42:49.8&-08:12:10.3&K7&440$\pm$30&23.3$\pm$0.1&130$\pm$20&0.35$\pm$0.02&0.5$\pm$0.1&-6$\pm$5   \\ 
313&05:40:25.7&-08:11:16.8&M3&400$\pm$40&1.40$\pm$0.10&7$\pm$1&-&-&-   \\ 
342&05:41:14.0&-08:07:57.4&M2&420$\pm$10&0.62$\pm$0.10&3.3$\pm$0.5&-&-&-   \\ 
378&05:41:30.6&-08:04:47.9&K7&420$\pm$20&29.1$\pm$0.1&150$\pm$20&0.12$\pm$0.04&1.0$\pm$0.5&70$\pm$70   \\ 
383&05:40:37.3&-08:04:03.0&F0&430$\pm$5&9.21$\pm$0.10&51$\pm$1&-&-&-   \\ 
387&05:40:41.0&-08:02:18.6&M0&470$\pm$90&2.38$\pm$0.10&16$\pm$6&-&-&-   \\ 
399&05:41:49.7&-08:00:32.1&K5&396$\pm$6&36.4$\pm$0.1&171$\pm$5&0.26$\pm$0.01&1.00$\pm$0.09&0$\pm$70   \\ 
400&05:41:41.7&-08:00:18.4&M0&400$\pm$200&2.14$\pm$0.10&11$\pm$10&-&-&-   \\ 
402&05:41:33.4&-07:59:56.2&M1.5&310$\pm$50&5.13$\pm$0.10&15$\pm$5&-&-&-   \\ 
403&05:41:54.6&-07:59:12.4&K3.5&410$\pm$30&1.68$\pm$0.10&9$\pm$1&-&-&-   \\ 
411&05:41:43.7&-07:58:22.4&M7.5&428$\pm$10$^a$&0.63$\pm$0.10&3.5$\pm$0.6&-&-&-   \\ 
428&05:40:27.8&-07:55:36.3&-&400$\pm$100&13.7$\pm$0.1&60$\pm$40&0.44$\pm$0.03&1.00$\pm$0.09&80$\pm$70   \\ 
429&05:40:24.9&-07:55:35.4&M3&390$\pm$20&7.19$\pm$0.10&34$\pm$3&-&-&-   \\ 
434&05:41:33.2&-07:55:02.1&K7&420$\pm$10&4.10$\pm$0.10&22$\pm$1&-&-&-   \\ 
435&05:41:21.4&-07:55:01.1&-&428$\pm$10$^a$&$<$0.48&$<$2.64&-&-&-   \\ 
463&05:41:25.9&-07:49:50.6&K7&430$\pm$40&0.99$\pm$0.10&6$\pm$1&-&-&-   \\ 
466&05:41:28.0&-07:49:22.4&M4&280$\pm$40&2.53$\pm$0.10&6$\pm$2&-&-&-   \\ 
468&05:40:17.1&-07:49:14.4&-&450$\pm$40&18.46$\pm$0.10&110$\pm$20&-&-&-   \\ 
471&05:40:18.5&-07:49:06.7&-&428$\pm$10$^a$&1.12$\pm$0.10&6.2$\pm$0.6&-&-&-   \\ 
474&05:40:43.6&-07:48:47.8&-&428$\pm$10$^a$&0.54$\pm$0.10&3.0$\pm$0.6&-&-&-   \\ 
476&05:40:59.9&-07:48:16.0&M3&430$\pm$70&1.91$\pm$0.10&11$\pm$4&-&-&-   \\ 
477&05:40:57.5&-07:48:08.8&M1.5&428$\pm$10$^a$&2.64$\pm$0.10&14.5$\pm$0.9&-&-&-   \\ 
483&05:41:05.5&-07:47:07.6&G1&400$\pm$30&9.2$\pm$0.1&44$\pm$6&0.42$\pm$0.04&0.8$\pm$0.2&10$\pm$20   \\ 
485&05:40:49.3&-07:46:32.5&M0&500$\pm$100&1.27$\pm$0.10&9$\pm$4&-&-&-   \\ 
487&05:41:04.6&-07:45:40.1&M0.5&428$\pm$10$^a$&2.92$\pm$0.10&16.1$\pm$0.9&-&-&-   \\ 
488&05:40:44.1&-07:45:09.7&-&410$\pm$50&5.85$\pm$0.10&30$\pm$7&-&-&-   \\ 
491&05:40:44.2&-07:44:43.5&-&428$\pm$10$^a$&8.84$\pm$0.10&49$\pm$2&-&-&-   \\ 
494&05:40:44.4&-07:44:16.7&-&428$\pm$10$^a$&$<$0.48&$<$2.64&-&-&-   \\ 
512&05:40:13.8&-07:32:16.1&M5.5&410$\pm$40&15.9$\pm$0.1&80$\pm$20&0.24$\pm$0.03&1.0$\pm$0.3&0$\pm$70   \\ 
525&05:40:44.7&-07:29:54.5&M1.5&450$\pm$50&11.13$\pm$0.10&70$\pm$20&-&-&-   \\ 
530&05:40:15.3&-07:28:46.8&-&400$\pm$50&15.9$\pm$0.1&80$\pm$20&0.22$\pm$0.04&1.0$\pm$0.2&-40$\pm$70   \\ 
546&05:39:49.1&-07:26:17.2&-&428$\pm$10$^a$&1.65$\pm$0.10&9.1$\pm$0.7&-&-&-   \\ 
553&05:39:40.5&-07:26:03.5&-&428$\pm$10$^a$&0.84$\pm$0.10&4.6$\pm$0.6&-&-&-   \\ 
556&05:40:20.4&-07:25:54.0&M1.5&410$\pm$10&8.86$\pm$0.10&45$\pm$3&-&-&-   \\ 
561&05:39:58.9&-07:25:33.5&-&428$\pm$10$^a$&11.96$\pm$0.10&66$\pm$3&-&-&-   \\ 
574&05:40:06.5&-07:23:58.7&M3&410$\pm$80&3.72$\pm$0.10&19$\pm$7&-&-&-   \\ 
582&05:39:32.3&-07:22:24.2&K3&340$\pm$30&$<$0.48&$<$1.66&-&-&-   \\ 
597&05:38:50.0&-07:20:18.5&-&384$\pm$9&2.75$\pm$0.10&12.2$\pm$0.7&-&-&-   \\ 
598&05:39:44.3&-07:20:10.5&-&428$\pm$10$^a$&1.06$\pm$0.10&5.8$\pm$0.6&-&-&-   \\ 
619&05:38:55.0&-07:11:53.7&-&800$\pm$900&1.97$\pm$0.10&40$\pm$90&-&-&-   \\ 
626&05:38:36.6&-07:11:00.2&M4&370$\pm$30&2.39$\pm$0.10&10$\pm$1&-&-&-   \\ 
633&05:38:17.4&-07:09:39.5&M3&390$\pm$20&1.19$\pm$0.10&5.3$\pm$0.7&-&-&-   \\ 
637&05:38:23.9&-07:07:38.9&K8&428$\pm$10$^a$&2.07$\pm$0.10&11.4$\pm$0.8&-&-&-   \\ 
641&05:38:13.4&-07:06:43.4&K4.5&370$\pm$60&1.68$\pm$0.10&7$\pm$2&-&-&-   \\ 
644&05:38:47.7&-07:06:14.8&M1.5&400$\pm$100&2.73$\pm$0.10&13$\pm$9&-&-&-   \\ 
645&05:38:41.5&-07:05:59.3&-&428$\pm$10$^a$&14.5$\pm$0.1&80$\pm$4&0.34$\pm$0.03&0.8$\pm$0.1&30$\pm$20   \\ 
653&05:38:09.1&-07:05:25.8&K2&428$\pm$10$^a$&0.85$\pm$0.10&4.7$\pm$0.6&-&-&-   \\ 
654&05:38:46.8&-07:05:09.1&M0.5&390$\pm$50&2.4$\pm$0.1&11$\pm$3&-&-&-   \\ 
663&05:38:44.0&-07:03:09.5&-&428$\pm$10$^a$&0.6$\pm$0.1&3.1$\pm$0.6&-&-&-   \\ 
666&05:38:44.8&-07:02:47.0&-&300$\pm$100&0.9$\pm$0.1&3$\pm$2&-&-&-   \\ 
673&05:38:04.8&-07:02:21.7&M5&340$\pm$40&$<$0.48&$<$1.64&-&-&-   \\ 
677&05:38:56.5&-07:01:55.4&-&400$\pm$200&4.1$\pm$0.1&20$\pm$20&-&-&-   \\ 
680&05:38:41.5&-07:01:52.5&M2&390$\pm$50&1.9$\pm$0.1&9$\pm$2&-&-&-   \\ 
689&05:39:01.2&-07:01:09.5&M3&420$\pm$50&3.1$\pm$0.1&16$\pm$4&-&-&-   \\ 
729&05:37:54.5&-06:57:31.1&K7&428$\pm$10$^a$&0.7$\pm$0.1&3.9$\pm$0.6&-&-&-   \\ 
734&05:37:51.7&-06:56:51.8&M2.4&428$\pm$10$^a$&0.6$\pm$0.1&3.4$\pm$0.6&-&-&-   \\ 
751&05:37:49.3&-06:51:37.4&M3.5&390$\pm$10&4.1$\pm$0.1&19$\pm$1&-&-&-   \\ 
755&05:37:44.5&-06:50:36.8&M1&390$\pm$8&1.4$\pm$0.1&6.6$\pm$0.5&-&-&-   \\ 
761&05:37:56.0&-06:48:54.9&M0&440$\pm$70&9.1$\pm$0.1&50$\pm$20&-&-&-   \\ 
762&05:36:08.3&-06:48:36.3&M2.5&390$\pm$10&20.6$\pm$0.1&95$\pm$6&0.28$\pm$0.03&0.9$\pm$0.1&-30$\pm$60   \\ 
792&05:37:58.8&-06:43:33.7&-&380$\pm$20&1.7$\pm$0.1&7.2$\pm$0.8&-&-&-   \\ 
798&05:36:30.2&-06:42:46.3&M1.5&430$\pm$40&0.6$\pm$0.1&3.5$\pm$0.9&-&-&-   \\ 
811&05:36:41.0&-06:41:17.8&M5&390$\pm$20&3.8$\pm$0.1&17$\pm$1&-&-&-   \\ 
818&05:37:32.4&-06:39:05.1&M3&356$\pm$8&0.6$\pm$0.1&2.4$\pm$0.4&-&-&-   \\ 
832&05:34:15.8&-06:36:04.5&G8&428$\pm$10$^a$&7.6$\pm$0.1&42$\pm$2&-&-&-   \\ 
847&05:37:00.1&-06:33:27.4&G3&428$\pm$10$^a$&0.9$\pm$0.1&5.0$\pm$0.6&-&-&-   \\ 
848&05:36:36.9&-06:33:24.2&K7.5&380$\pm$10&5.8$\pm$0.1&25$\pm$2&-&-&-   \\ 
853&05:34:06.9&-06:32:08.0&M0&390$\pm$20&$<$0.48&$<$2.23&-&-&-   \\ 
874&05:35:18.9&-06:27:25.3&M4&362$\pm$10&1.5$\pm$0.1&5.7$\pm$0.5&-&-&-   \\ 
887&05:35:45.9&-06:25:59.1&G0.5&400$\pm$20&$<$0.48&$<$2.29&-&-&-   \\ 
914&05:36:12.6&-06:23:39.4&M5&360$\pm$10&1.5$\pm$0.1&5.9$\pm$0.6&-&-&-   \\ 
920&05:35:37.3&-06:23:26.7&M2.5&379$\pm$10&1.6$\pm$0.1&7.1$\pm$0.6&-&-&-   \\ 
926&05:36:30.1&-06:23:10.1&K6.5&428$\pm$10$^a$&6.0$\pm$0.1&33$\pm$2&-&-&-   \\ 
930&05:35:41.0&-06:22:45.4&M0&428$\pm$10$^a$&3.2$\pm$0.1&17.7$\pm$1.0&-&-&-   \\ 
971&05:36:32.3&-06:19:19.9&G8&387$\pm$5&6.5$\pm$0.1&29.1$\pm$0.8&-&-&-   \\ 
994&05:35:14.6&-06:15:12.5&M1&390$\pm$20&1.8$\pm$0.1&8.1$\pm$0.8&-&-&-   \\ 
1001&05:35:27.9&-06:14:15.0&K8&400$\pm$10&6.8$\pm$0.1&32$\pm$2&-&-&-   \\ 
1006&05:36:40.4&-06:13:33.3&M0.5&385$\pm$8&5.0$\pm$0.1&22.1$\pm$1.0&-&-&-   \\ 
1007&05:36:14.7&-06:13:16.9&M1.5&369$\pm$7&10.2$\pm$0.1&42$\pm$2&0.27$\pm$0.07&1.0$\pm$0.3&-80$\pm$70   \\ 
1011&05:35:48.9&-06:12:07.7&K4&399$\pm$5&10.5$\pm$0.1&50$\pm$1&-&-&-   \\ 
1020&05:36:40.8&-06:11:08.2&K2&410$\pm$20&1.0$\pm$0.1&5.2$\pm$0.7&-&-&-   \\ 
1039&05:36:26.1&-06:08:03.7&M0&380$\pm$7&20.7$\pm$0.1&90$\pm$3&0.30$\pm$0.03&0.9$\pm$0.1&-70$\pm$40   \\ 
1086&05:34:58.5&-06:00:00.5&M0&428$\pm$10$^a$&17.9$\pm$0.1&99$\pm$5&0.25$\pm$0.03&0.6$\pm$0.3&10$\pm$10   \\
\enddata
\tablecomments{$^a$ Objects either without distances in the \cite{bailer-jones20} catalogue or with a Renormalized Unit Weight Error greater than 1.4; these are listed with the distance of 428$\pm$10 pc from \cite{kounkel17}. Spectral types were collected from the literature and are listed with references in \cite{grant18}. Dust masses are determined using the distances listed for each target. Columns 8-10 are from deconvolved fits to the continuum images (see Section~\ref{subsec: continuum results}).}
%%% From Table1.txt made from Table1.py in /Cycle7/Scripts/
\end{deluxetable*}
\clearpage

\begin{deluxetable*}{cccccccc}[ht!]
\tablecaption{Additional Detected Objects\label{tab: extras table}}
\tablehead{ \colhead{Num.}& 
\colhead{{[}MGM2012{]} ID} & \colhead{RA} & \colhead{Dec} & \colhead{F$_{1.33\ mm}$} & \colhead{Major} & \colhead{Ratio} & \colhead{PA} \\
\colhead{} & \colhead{} & \colhead{(J2000)} & \colhead{(J2000)} & \colhead{(mJy)}  & \colhead{(arcsec)} & \colhead{} & \colhead{($^\circ$)}
}
\startdata
1&&05:42:50.3&-08.39.54.2&3.25$\pm$0.10&-&-&-   \\ 
2&379&05:41:30.3&-08.04.42.4&1.2$\pm$0.2$^{a}$&-&-&-   \\ 
3&&05:40:42.8&-07.48.40.3&0.7$\pm$0.1$^{a}$&-&-&-   \\ 
4&591&05:38:50.5&-07.20.29.8&1.05$\pm$0.10&-&-&-   \\ 
5&596&05:39:43.4&-07.20.21.0&2.11$\pm$0.10&-&-&-   \\ 
6&&05:38:04.6&-07.02.23.6&6.6$\pm$0.1&0.32$\pm$0.07&1.0$\pm$0.4&0$\pm$70   \\ 
7&&05:38:42.1&-07.01.52.8&1.5$\pm$0.1&-&-&-   \\ 
8&995&05:35:14.6&-06.15.07.4&0.6$\pm$0.1&-&-&-   \\ 
\enddata
\tablecomments{$^a$ Objects where the \texttt{uvmodelfit} point source and Gaussian fits did not converge in the same location, making these objects very marginally detected. For these objects we report fluxes from \texttt{imfit} instead.  }
\end{deluxetable*}

\section{Results}\label{sec: results}

\subsection{1.33 mm Continuum Results}\label{subsec: continuum results}
Here, we discuss the continuum observations for our sample. First, we present the criteria for detection in the continuum and the continuum fluxes. Second, we compute the dust masses of the sources in our sample. Finally, we compare the dust mass distribution of our sample to Class II surveys in other star-forming regions and to the VANDAM protostellar disks in L1641. 

\subsubsection{Continuum Detections and Fluxes}\label{subsubsec: continuum detections}

\begin{figure*}
    \centering
    \includegraphics[scale=0.73]{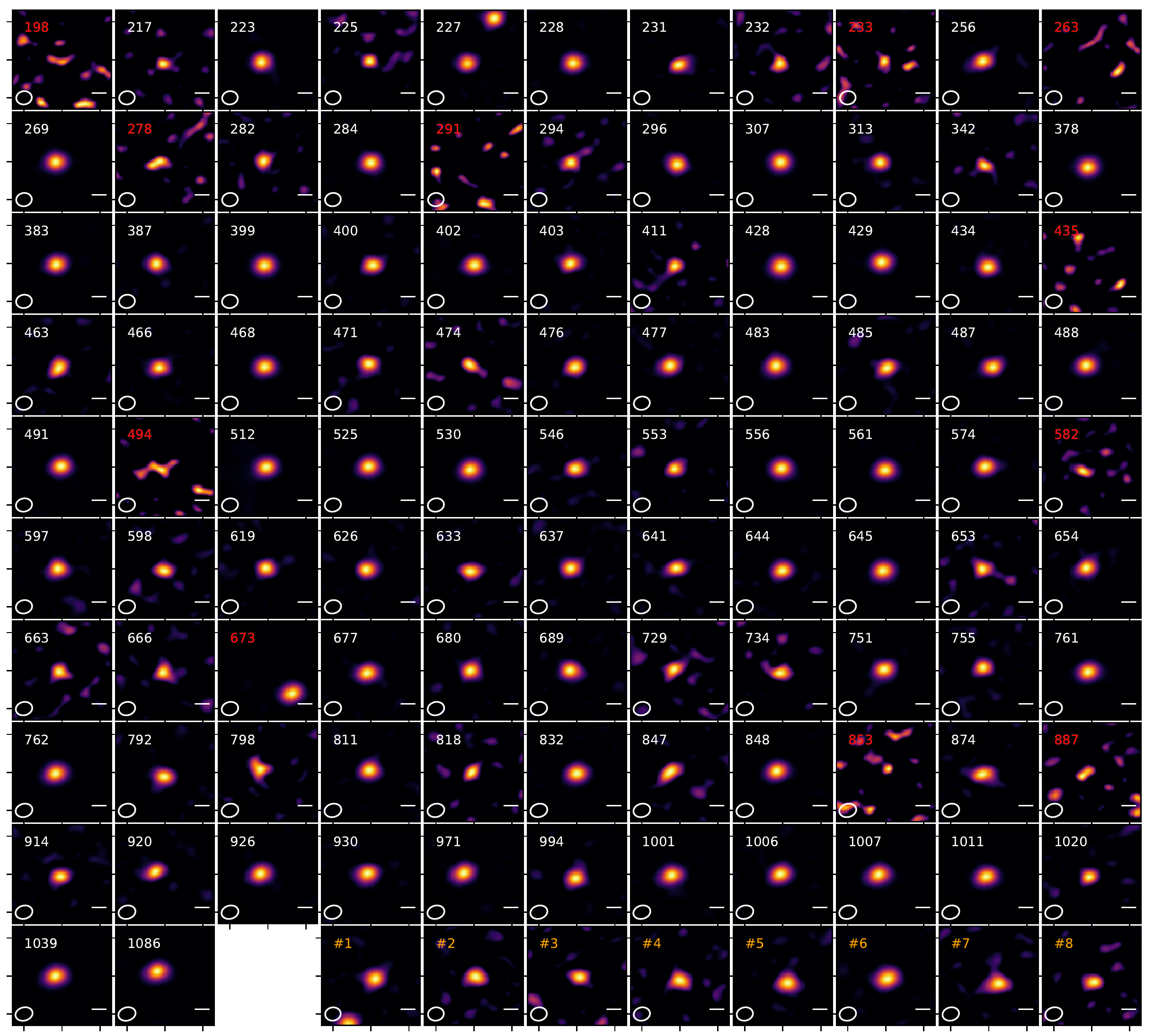}
    \caption{Continuum images at 1.33 mm for the 101 disks in our ALMA sample. The beam for each observation is shown along with a bar of 500 AU (assuming 428 pc distance). Each image is 8''$\times$8'' and the tick marks are 3'' apart. The minimum color value is set to 0.12 mJy beam$^{-1}$, the typical rms of our observations. The IDs correspond to the [MGM2012] numbers and objects that are considered non-detections have their IDs written in red. The additional objects that are detected in the continuum are shown in the bottom right with orange labels. The numbers correspond to the numbers in Table~\ref{tab: extras table}. }
    \label{fig: continuum images}
    % from FitsImages.py
\end{figure*}

We use the CASA \texttt{imfit}, \texttt{imstat} and  \texttt{uvmodelfit}  tasks to determine continuum properties for each target in our sample. For \texttt{imfit} and \texttt{imstat}, we use a box region located at the image center (except in the case of [MGM2012] 673 which is slightly off-center to avoid a nearby source, \#6 in Table~\ref{tab: extras table}) with a width and height of 3.8''x3.8'' or 1630x1630 AU at a distance of 428 pc. We also use the average rms value of the observations of 0.12 mJy beam$^{-1}$ as an input into \texttt{imfit}. If the Gaussian fit converges, then \texttt{imfit} provides, among other properties, the flux of the Gaussian fit and the peak flux. We use \texttt{imstat} to determine the maximum flux in the same boxed region. We consider objects detected if the maximum flux value found with \texttt{imstat} is greater than four times the average rms of the observations. In our sample, 89 of 101 disks are detected given this criterion (88\%). We use \texttt{uvmodelfit} point source models to determine the source fluxes and off-sets from phase center. Additionally, we fit the sources with a Gaussian in \texttt{uvmodelfit} which also provides the FWHM along the major axis, the aspect ratio between the major and minor axes, and the position angle. The \texttt{uvmodelfit} continuum fluxes are given in Table~\ref{tab: observations table} and used for the rest of the analysis. The typical $\chi^{2}$ of the \texttt{uvmodelfit} Gaussian and point source fits is 1.302. The flux uncertainties have been multiplied by the factor needed to reduce the $\chi^{2}$ value to 1. We show the continuum images in Figure~\ref{fig: continuum images}. For the additional sources detected in our observations that are not in our sample (Table~\ref{tab: extras table}), we performed the same steps and provide the fluxes for these objects in Table~\ref{tab: extras table} and show their continuum images in the bottom row of Figure~\ref{fig: continuum images}. For two of the additionally detected objects, the \texttt{uvmodelfit} point source and Gaussian fits converge at slightly different locations, likely due to the low signal-to-noise, therefore we report the fluxes from \texttt{imfit}.

As stated in Section~\ref{sec: observations}, the beam size of our observations has a major axis of $\sim$1.2-1.4 arcseconds and a minor axis of $\sim$1.04-1.05 arcseconds. We use the criteria of \cite{cieza19} to determine what objects are suitable for radii determinations. Specifically, we determine an object to be resolved if it: 1) has an S/N$>$30 (where S/N is the maximum pixel value in the source box from \texttt{imstat} divided by the average rms), and 2) has a Gaussian flux density from \texttt{uvmodelfit} that is greater than the \texttt{uvmodelfit} point source flux plus 3$\times$ the average rms. Doing this, we find that 13 targets are suitable for radial fits and we consider these objects to be resolved. Among the 13 disks that are resolved, three are transitional disks ([MGM2012] 269, 307, and 1086). We do not resolve cavities in any of the transitional disks, which is expected given our low angular resolution. The FWHM value for the major axis, the ratio between the axes, and position angle (PA) from \texttt{uvmodelfit} are provided for the resolved targets in Table~\ref{tab: observations table}. The uncertainties on these values have been treated the same as for the flux uncertainties discussed above. The \texttt{uvmodelfit} documentation cautions that the size uncertainties for the Gaussian fits may be underestimated. We suggest interpreting the size information with caution, especially given our limited spatial resolution. We note that only one of the additional targets detected in the ALMA fields (Table~\ref{tab: extras table}) meet the \cite{cieza19} criteria for reporting the Gaussian fit values.

\begin{figure}[b!]
    \centering
    \includegraphics[scale=0.55]{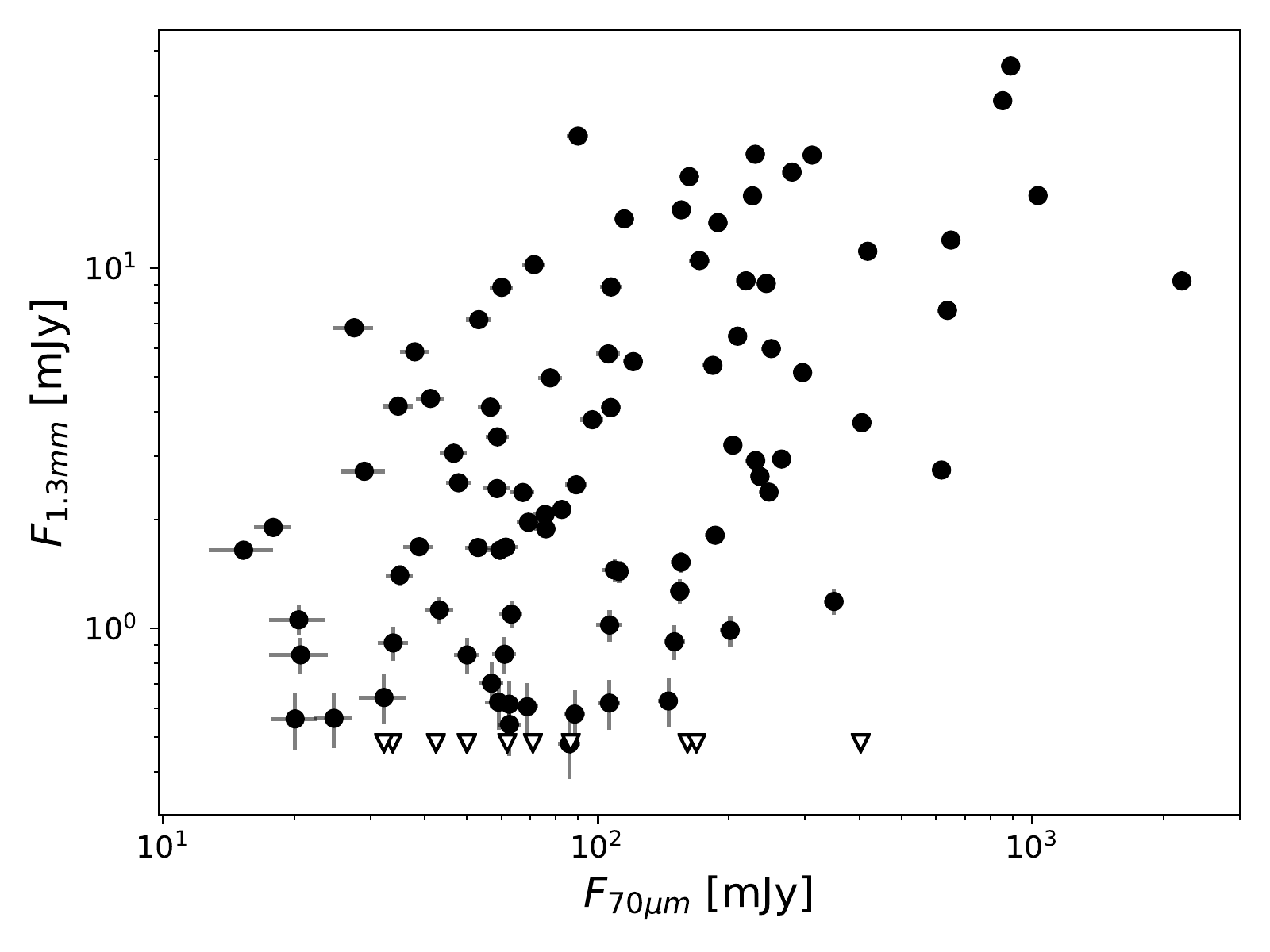}
    \caption{ALMA 1.33 mm continuum flux vs. \textit{Herschel} PACS 70 \mic\ flux. Upper limits are shown as the open downward-facing triangles. There is a moderate Pearson correlation coefficient between the two of 0.44.}
    \label{fig: 70mic vs 1.3mm}
    % from Analysis.py
\end{figure}

\begin{figure*}[t!]
    \plottwo{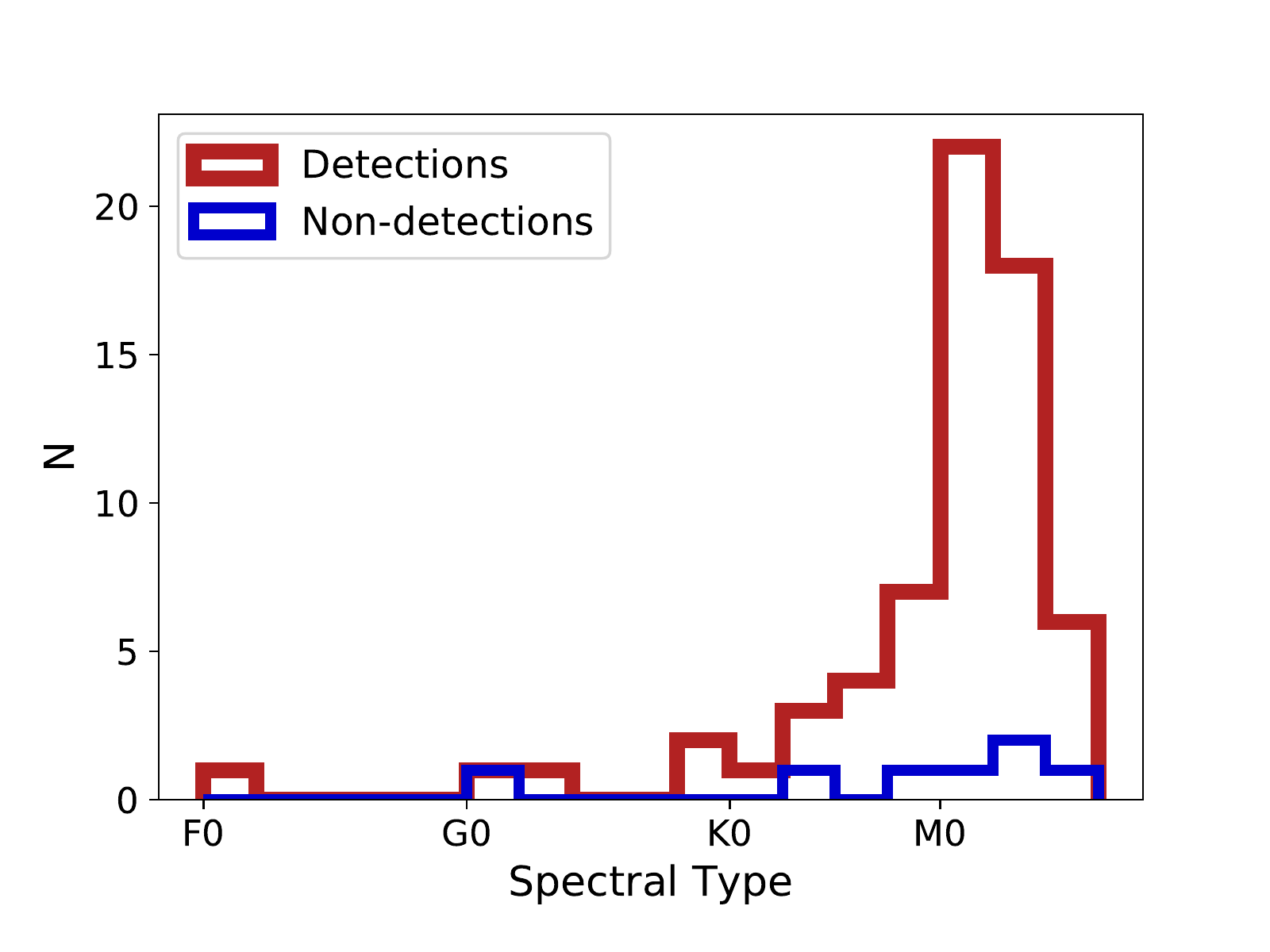}{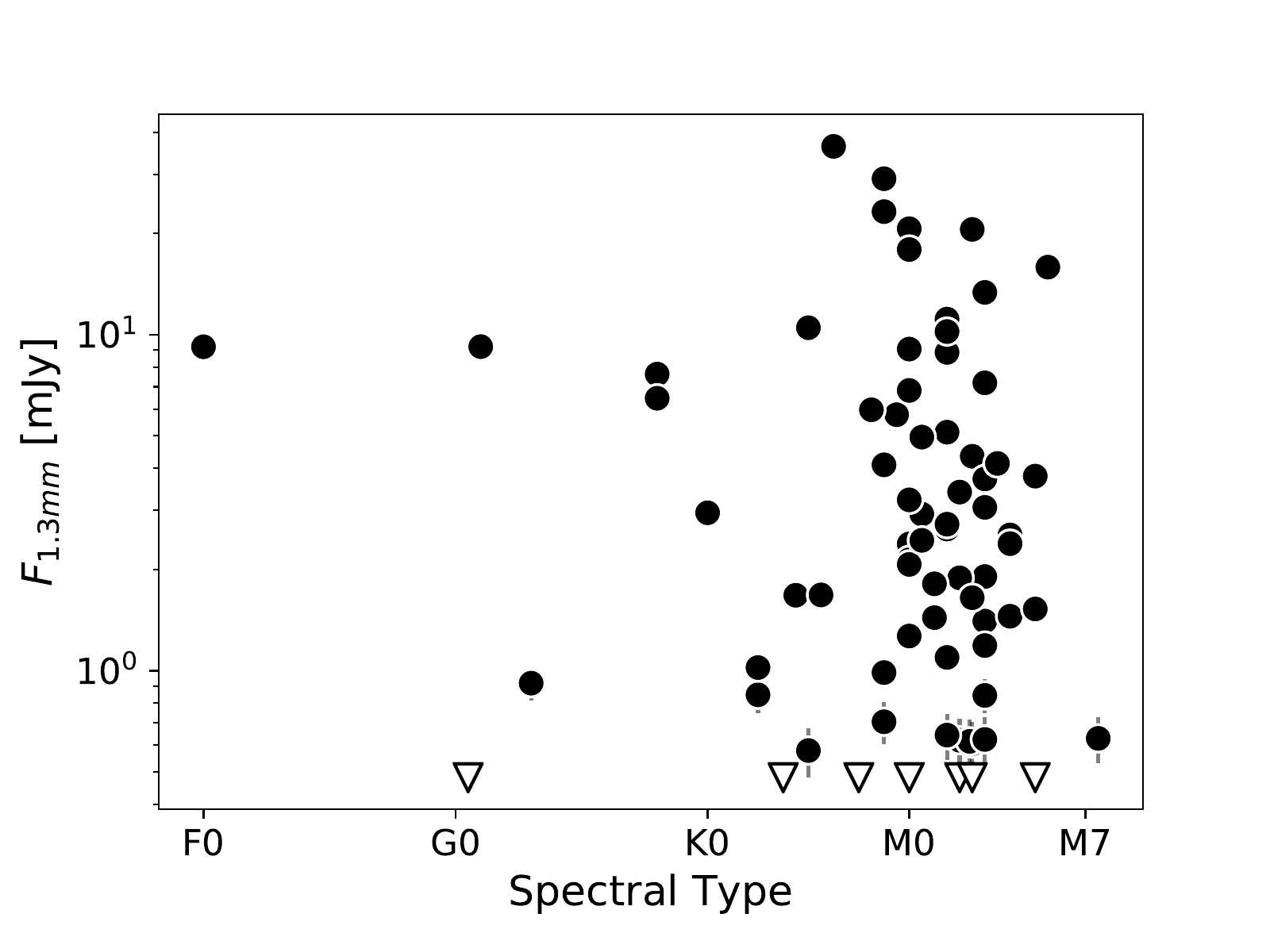}
    \caption{Left: The distribution of spectral types for the objects detected with ALMA (red) and not detected (blue). Right: The 1.33 mm ALMA flux as a function of spectral type for objects with known spectral types. Upper limits are shown as the open downward-facing triangles. \label{fig: SpT flux}}
    \end{figure*}

Figure~\ref{fig: 70mic vs 1.3mm} shows the 1.33 mm ALMA flux vs. the 70 \mic\ \textit{Herschel} flux. There is a moderate correlation between the two with a Pearson coefficient of 0.45 using the \texttt{scipy.stats.pearsonr} function. Figure~\ref{fig: SpT flux} shows the distribution of spectral types for continuum detections and non-detections (left). Spectral types are compiled from the literature, with over 75\% coming from \cite{hsu12}, and are provided in Table~\ref{tab: observations table} (see \citealt{grant18} for references). The most common spectral type in this sample, M1 ($T_{eff}=3680$, \citealt{pecaut-mamajek13}), corresponds to a $\sim$0.4-0.5 \Msun\ pre-main-sequence star at 1 Myr \citep{baraffe15}. The L1641 spectral type distribution in \cite{hsu12} peaks at M4 with later spectral types incomplete due to extinction. \cite{fang13} finds a similar peak in the spectral types for L1641. Therefore, we are biased towards slightly earlier spectral types, and thus slightly higher stellar masses. Figure~\ref{fig: SpT flux} also shows the flux as a function of the spectral type showing that there is no clear trend with spectral type in the L1641 ALMA observations (right).

\subsubsection{Dust Masses}\label{subsubsec: dust masses}
\begin{figure*}[t!]
    \centering
    \includegraphics[scale=0.56]{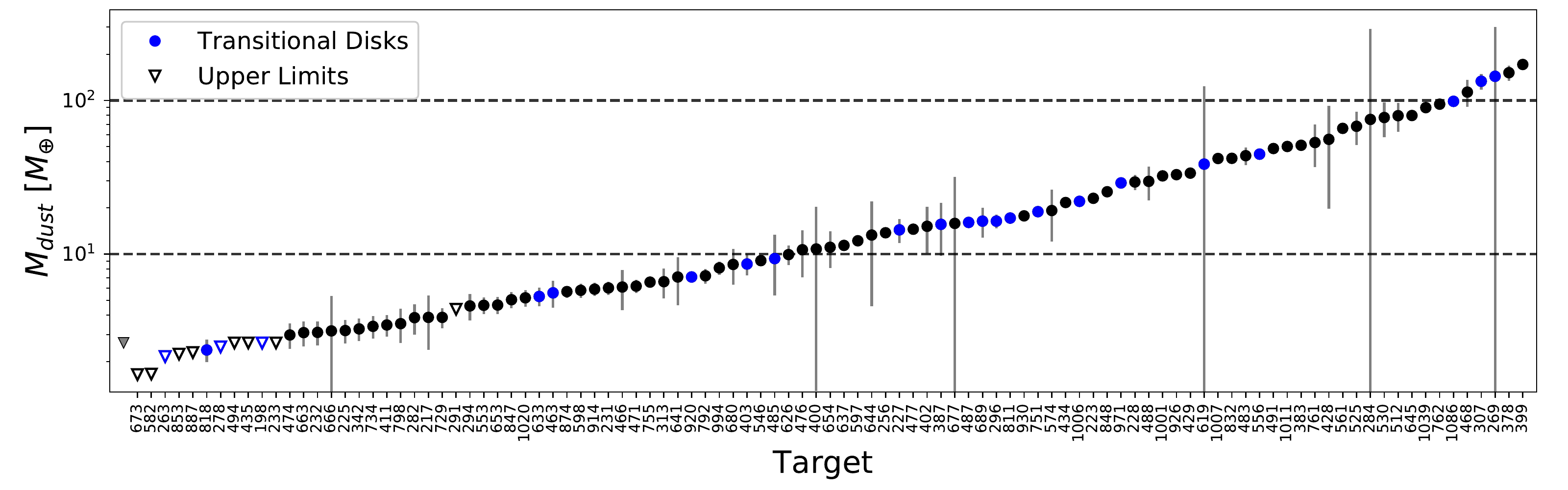}
    \caption{The dust mass in Earth masses for each target. Transitional disks from \cite{grant18} are shown in blue. Upper limits are shown as the open downward-facing triangles. The error does not include absolute flux calibration error. We use individual distances to each object as discussed in Section~\ref{subsubsec: dust masses}. The gray triangle to the left is the upper limit using four times the average rms in the observations and a distance of 428 pc. }
    \label{fig: targ mdust}
    % from Analysis.py
\end{figure*}

We determine the dust mass in the disk using the following relation that is appropriate for optically thin emission:

\begin{equation}\label{eq 1}
    M_{dust} = \frac{d^{2}F_{\nu}}{\kappa_{\nu}B_{\nu}(T_{dust})},
\end{equation}
where $d$ is the distance to the source, $F_{\nu}$ is the flux density of the source, $\kappa_{\nu}$ is the dust opacity given by $\kappa_{\nu} = (\nu/100$ GHz) cm$^2$ g$^{-1}$ at 1.33 mm (2.254 cm$^2$ g$^{-1}$, \citealt{beckwith90} with a spectral index of 1), and $B_{\nu}(T_{dust})$ is the Planck function given a uniform dust temperature of 20 K. We use Gaia distances from \cite{bailer-jones20} for all sources where distances are available. For objects without distances in \cite{bailer-jones20}, or where the Renormalized Unit Weight Error (RUWE) is greater than 1.4, we adopt the distance to L1641 from \cite{kounkel17} of 428$\pm$10 pc. For our undetected sources, we take an upper flux limit of 4 times the average rms (0.12 mJy beam$^{-1}$) which we convert to a dust mass given the object's distance. At 428 pc this corresponds to 2.6 \Mearth\ using Equation~\ref{eq 1}. The median dust mass in the L1641 sample is 11.1$^{+32.9}_{-4.6}$ \Mearth, including upper limits. The errors are the first and third quartiles of the distribution. Figure~\ref{fig: targ mdust} shows the dust mass for each target.

\cite{grant18} used the spectral energy distributions of the L1641 targets to determine whether they were transitional (see \citealt{grant18} for full details, but generally, the transitional disks are identified by a lack of emission in the \textit{Spitzer} IRAC bands). Of the 23 transitional disks in our sample observed by ALMA, 20 are detected. The transitional disks have a median dust mass of 16.1 \Mearth, including upper limits. The median of the transitional dust mass is larger than for the sample as a whole. However, the samples are statistically similar, with a two-sample Kolmogorov-Smirnov p-value of 0.72.

\subsubsection{Dust Mass Comparison}\label{subsubsec: dust mass comparison}

In order to compare L1641 to targets of different evolutionary states and regions, we gather dust masses for the L1641 protostars in the VANDAM survey \citep{tobin20}, as well as Class II objects in several other star-forming regions. In \cite{tobin20} the dust masses for the protostellar disks are determined primarily using $\kappa_{\nu}$ = 1.84 and a dust temperature of 43 K that is scaled based on the protostar bolometric luminosity. To compare to the Class II sample in this work, we redetermine the dust masses using a dust temperature of 20 K and a dust opacity of $\kappa_{\nu}$ = $\nu$/100 GHz, corresponding to 3.45 cm$^2$ g$^{-1}$ at the observed wavelength of 0.87 mm. We note that \cite{tobin20} analyzed L1641 and the Southern ISF together as one population, however, we only analyze protostellar disks below -6$^{\circ}$ to match our sample. We also determine the dust masses of other regions surveyed in the millimeter in the same way as our ALMA sample. These regions include ONC (1 Myr, 400 pc, \citealt{eisner18}), OMC1 (1 Myr, 400 pc, \citealt{eisner16}), OMC2 (1 Myr, 414 pc, \citealt{vanTerwisga19}), Lupus (1-3 Myr, 150-200 pc, \citealt{ansdell16}), Corona Australis (1-3 Myr, 160 pc, \citealt{cazzoletti19}), Ophiuchus (1 Myr, 140 pc, \citealt{cieza19, williams19}), Chamaeleon I (2 Myr, 160 pc, \citealt{pascucci16}), Taurus (2 Myr, 140 pc, \citealt{andrews13}), IC 348 (2-3 Myr, 310 pc, \citealt{ruizrodriguez18}), $\sigma$ Orionis (3-5 Myr, 385 pc, \citealt{ansdell17}), and Upper Scorpius (5-11 Myr, 145 pc, \citealt{barenfeld16}).

\begin{figure*}
    \centering
    \includegraphics[scale=0.5]{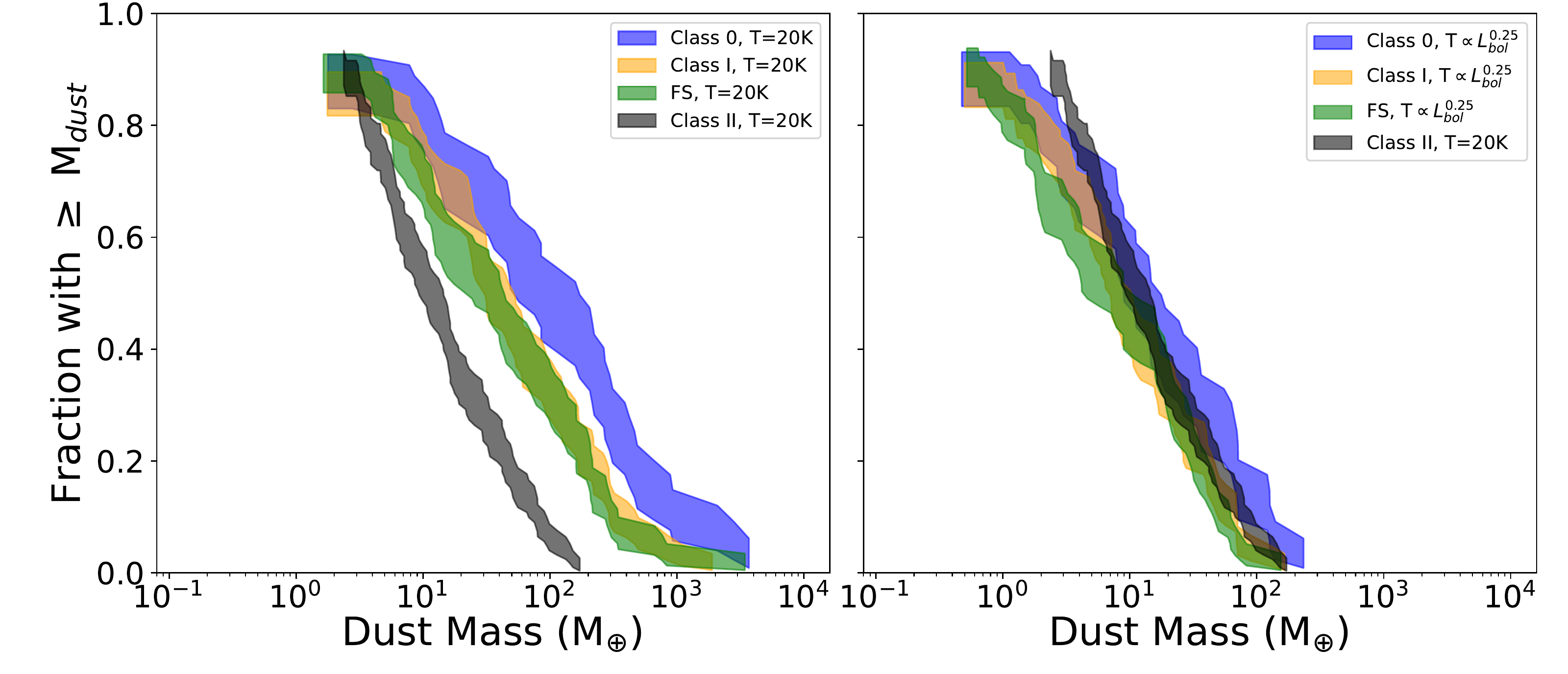}
    \caption{The cumulative dust mass distribution for objects in L1641. Left: Our L1641 Class II sample is shown in gray, the VANDAM L1641 Class 0, I, and Flat Spectrum sources are shown in blue, orange, and green, respectively. The VANDAM dust masses have been calculated using $T_{dust}$=20 K, $\kappa_{\nu}$=3.45 cm$^{2}$ g$^{-1}$. Right: Same as the left, but the VANDAM dust masses are calculated with a variable dust temperature, dependent on the protostellar bolometric luminosity (see \citealt{tobin20}).}
    \label{fig: Mdust cum L1641 classes}
    % from Analysis.py
\end{figure*}

To compare our L1641 sample to other regions/surveys, we make dust mass cumulative distributions using a Kaplan-Meier estimator from the \texttt{lifelines} package \citep{lifelines}\footnote{https://zenodo.org/record/3629409\#.XkMMaBNKhsM}. This estimator takes the non-detection upper limits into account for each sample with the \texttt{fit\_left\_censoring} function for left-censored data. In Figure~\ref{fig: Mdust cum L1641 classes} we show our Class II sample compared to the VANDAM L1641 protostellar disks, separated by evolutionary stage, with both the recalculated dust masses ($T_{dust}$=20 K, $\kappa_{\nu}$=3.45 cm$^{2}$ g$^{-1}$, assuming a spectral index of 1 for our adopted dust opacity) and those with a dust temperature that depends on each source's bolometric luminosity, as in \cite{tobin20}.  The dust masses are lower when the dust temperature is varied with the bolometric luminosity. The uniform temperature increases the dust mass by negating the inverse proportionality of dust mass and temperature. The distribution shifts to lower dust masses with increasing evolutionary classification from Class 0 to Flat Spectrum, regardless of the dust mass input parameters, although the range of protostellar disk masses is more narrow when the dust temperature is varied depending on bolometric luminosity.

\begin{figure*}
    \centering
    \includegraphics[scale=0.5]{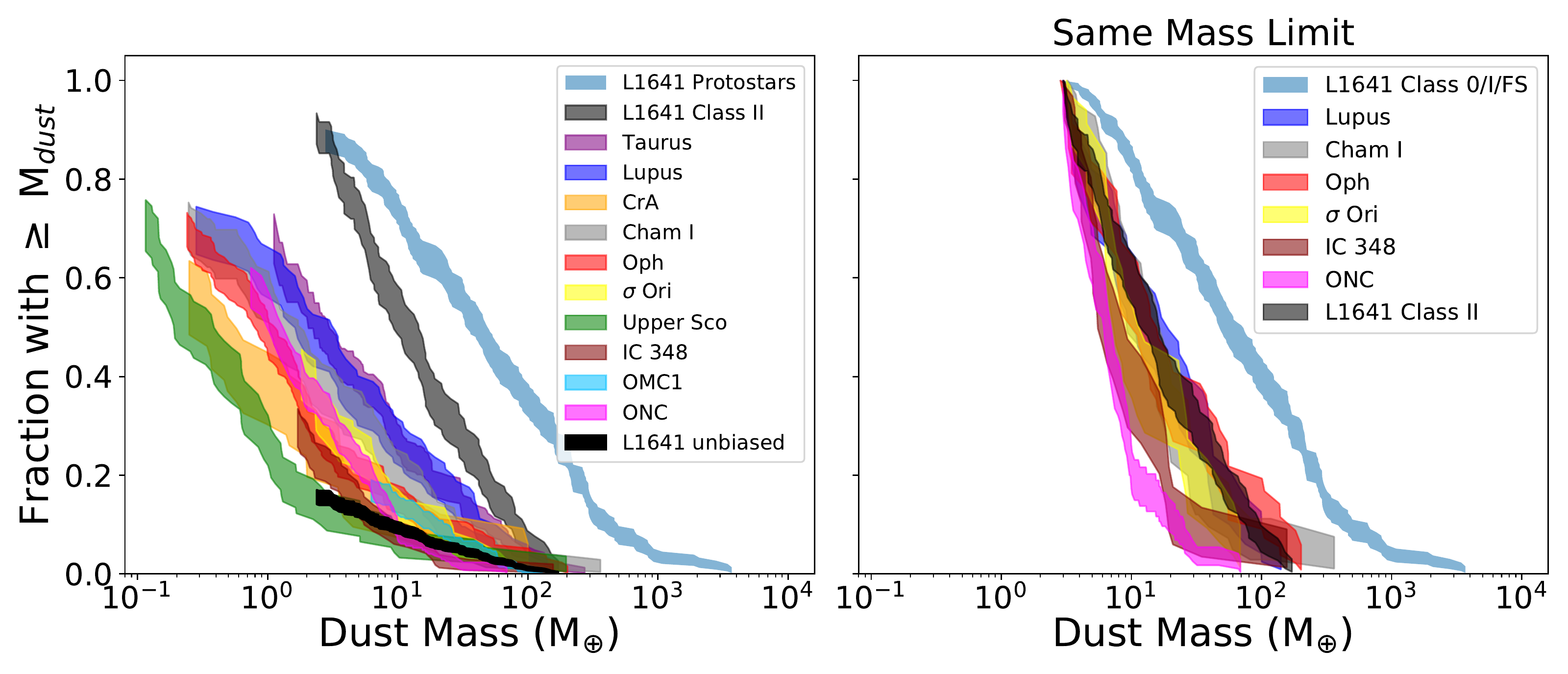}
    \caption{Left: The cumulative distribution for dust masses. There are no cuts due to the stellar population or sensitivities of the different surveys. The dark gray is the L1641 Class II sample, the black is the sample if we correct for the \textit{Herschel} bias as discussed in Section~\ref{subsubsec: dust mass comparison}. Right: The cumulative distribution for dust masses for detected disks with dust masses greater than the minimum detected mass in the VANDAM survey of protostars in L1641 (2.78 \Mearth). The comparison regions/surveys include, in order of increasing age, L1641 0/I/FS \citep{tobin20}, ONC \citep{eisner18}, OMC1 \citep{eisner16}, Lupus \citep{ansdell16}, Corona Australis \citep{cazzoletti19}, Ophiuchus \citep{williams19, cieza19}, Chamaeleon I \citep{pascucci16}, Taurus \citep{andrews13}, IC 348 \citep{ruizrodriguez18}, $\sigma$ Orionis \citep{ansdell17}, and Upper Scorpius \citep{barenfeld16}.  \label{fig: Mdust cum and VANDAM limit}}
    % from Analysis.py
\end{figure*}

Figure~\ref{fig: Mdust cum and VANDAM limit} (left) shows the cumulative dust mass distributions for some of the comparison regions and surveys. The height of the distribution corresponds to the detection rate of the survey. Some comparison surveys give very low or negative dust masses for some upper limit targets, thus we cut the distribution at the lowest detected dust mass for readability. We note that for this plot, the L1641 Class 0, I, and Flat Spectrum sources are combined into the L1641 Protostars distribution. The oldest star-forming region that we compare to, Upper Scorpius, is the lowest mass distribution while the highest mass distribution is the young VANDAM protostellar disks in L1641. Our L1641 Class II sample is quite distinct from the rest of the Class II surveys. This is due to the high detection rate in this survey and the massive disks that we find. Both of these factors are affected by the bias in our sample selection. 

In Figure~\ref{fig: Mdust cum and VANDAM limit} (left) we also show the distribution when we correct for the \textit{Herschel} bias discussed in Section~\ref{sec: sample} (L1641 unbiased, black). Specifically, we take the additional 480 Class IIs that fell into the HOPS fields (but were not detected at 70 \mic\ and not included in our ALMA sample) and turn them into upper limits in the \texttt{lifelines} function. The moderate correlation between 70 \mic\ flux and the flux at 1.33 mm, shown in Figure~\ref{fig: 70mic vs 1.3mm}, indicates that objects that were not detected by \textit{Herschel} would have lower dust masses than our detection threshold, and therefore, would not be detected in our ALMA observations. However, as the correlation between the fluxes at 70 \mic\ and 1.33 mm is moderate, some of the sources not detected by \textit{Herschel} may be able to be detected by ALMA given the same flux limit, making this a conservative estimate. Making these additional 480 objects into upper limits makes our detection rate go from 87\% to 14\%. This shows the importance of taking the completeness of a sample and sample biases into account. This is an estimate and a complete, unbiased survey of this region will be necessary to determine where the region as a whole lies on this distribution.  

Not all surveys used for comparison have the same sensitivity limits and because many of the comparison regions are closer than L1641, this impacts what objects are detectable. To determine how this affects the results, we determine the minimum dust mass detected by each comparison survey. By choosing a minimum mass as a lower limit, we can compare detected sources above that limit and reduce differences due to sensitivity, although this does not take into account biases in the target selection. The cumulative distribution is shown in Figure~\ref{fig: Mdust cum and VANDAM limit} (right) when the lower limit is taken to be the 2.8 \Mearth\ limit of the L1641 protostars in the VANDAM survey. This comparison shows that L1641 lies with the other regions in a tightly packed distribution, with ONC disks at lower values than the other surveys.

\begin{figure*}
    \centering
    \includegraphics[scale=0.6]{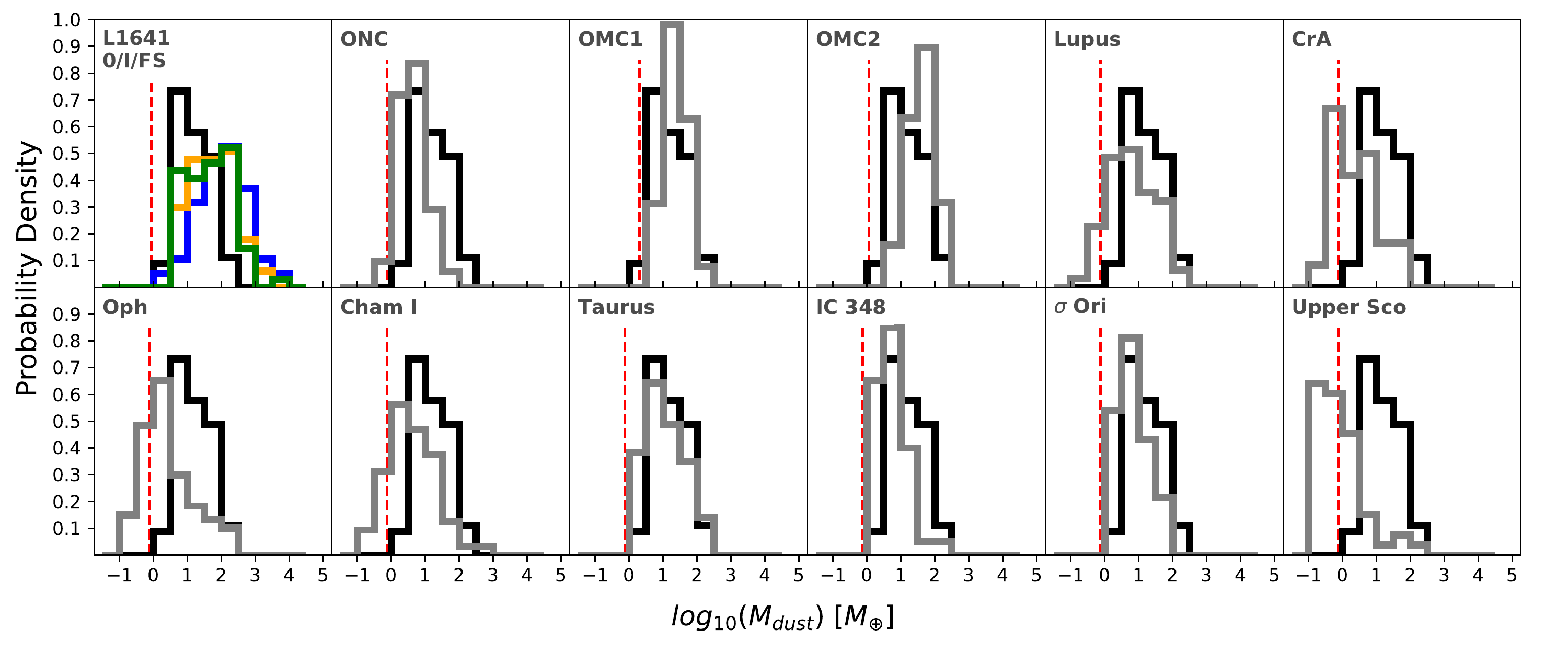}
    \caption{Here, we show the normalized dust mass distributions for the objects of our L1641 Class II sample (black) compared to other regions and populations surveyed with ALMA (gray). The protostellar disks in the top left panel are broken into Class 0 (blue), Class I (orange), and Flat Spectrum (green). Age of the comparison sample increases to the right. Only detected objects are included here. These samples all have different sensitivities. We determine the lowest detected mass in each sample. The red-dashed line is the higher of the lower limits of the comparison survey and our survey. Only the distributions to the right of this line are comparable. The comparison regions/surveys are the same as in Figure~\ref{fig: Mdust cum and VANDAM limit}.}
    \label{fig: Mdust hist}
    % from Analysis.py
\end{figure*}

In Figure~\ref{fig: Mdust hist} we show the distribution of dust masses for the L1641 Class II sample compared to other surveyed regions. We include only the detected objects here. In general, we find that the dust mass distribution shifts to lower masses with increasing age of the sample. The L1641 Class II sample is biased towards more massive disks, indicating that a complete sample of the Class IIs in L1641 may have a distribution also shifted to lower dust masses and thus the comparisons here are limited by the detection limits in each sample. We determine the lowest detected mass in each comparison survey. The red-dashed lines in Figure~\ref{fig: Mdust hist} correspond to the lowest detected dust mass in the distribution with the lower sensitivity. Below this limit, the surveys are not comparable. We find that the dust mass distribution for the L1641 Class IIs is comparable to the distributions of OMC1, OMC2, and Taurus, while the peaks of distributions of Upper Scorpius and Corona Australis are shifted to an order of magnitude or more lower in mass.

%\FloatBarrier 
%\startlongtable
\begin{deluxetable*}{cccc}
\tablecaption{CO Gas Properties \label{tab: gas properties}}
% \tabletypesize{\scriptsize}
\tablehead{
\colhead{{[}MGM2012{]} ID} & \colhead{F$_{12CO}$} & \colhead{F$_{13CO}$} & \colhead{F$_{C18O}$}  \\
\colhead{} & \colhead{(Jy km s$^{-1}$)} & \colhead{(Jy km s$^{-1}$)} & \colhead{(Jy km s$^{-1}$)}
}
\startdata
198 & -& 0.74$\pm$0.03& -   \\ 
227 & 0.15$\pm$0.03& -& -   \\ 
269 & 0.61$\pm$0.03& 0.09$\pm$0.02& -   \\ 
284 & 0.82$\pm$0.02& -& -   \\ 
291 & 1.19$\pm$0.03& -& -   \\ 
307 & 0.37$\pm$0.04& 0.21$\pm$0.04& -   \\ 
378 & 2.93$\pm$0.04& 0.62$\pm$0.04& 0.15$\pm$0.03   \\ 
399 & 1.32$\pm$0.06& 0.16$\pm$0.02& -   \\ 
400 & 0.38$\pm$0.03& -& -   \\ 
402 & 0.52$\pm$0.04& -& -   \\ 
428 & -& 0.18$\pm$0.02& -   \\ 
429 & 0.60$\pm$0.04& -& -   \\ 
435 & 0.65$\pm$0.05& -& -   \\ 
477 & 0.26$\pm$0.03& -& -   \\ 
483 & 0.18$\pm$0.02& -& -   \\ 
487 & 0.2$\pm$0.2& -& -   \\ 
488 & 0.10$\pm$0.02& -& -   \\ 
512 & -& 0.95$\pm$0.03& 0.45$\pm$0.03   \\ 
525 & 0.67$\pm$0.04& 0.26$\pm$0.03& -   \\ 
530 & 1.58$\pm$0.04& -& -   \\ 
546 & 0.30$\pm$0.03& -& -   \\ 
556 & 0.29$\pm$0.03& 0.24$\pm$0.03& 0.08$\pm$0.02   \\ 
561 & 2.29$\pm$0.03& -& 0.13$\pm$0.02   \\ 
633 & 0.23$\pm$0.02& -& -   \\ 
645 & 1.66$\pm$0.03& 0.26$\pm$0.03& -   \\ 
761 & 1.01$\pm$0.04& -& -   \\ 
762 & 1.57$\pm$0.03& 0.37$\pm$0.03& -   \\ 
848 & 0.45$\pm$0.04& -& -   \\ 
971 & 0.41$\pm$0.04& 0.20$\pm$0.03& -   \\ 
1001 & 0.83$\pm$0.05& -& -   \\ 
1006 & 0.18$\pm$0.03& -& -   \\ 
1007 & 0.28$\pm$0.03& -& -   \\ 
1039 & 0.93$\pm$0.04& 0.19$\pm$0.02& -   \\ 
1086 & 0.26$\pm$0.03& -& -   \\  
\enddata
% \tablecomments{Gas caption}
%%% From Analysis_gas.py
\end{deluxetable*}
% \clearpage

\subsection{CO Gas Detections and Fluxes}

Gas fluxes in the $^{12}$CO, $^{13}$CO, and C$^{18}$O line data were obtained via a Gaussian fit to a 3''$\times$3'' circular region centered on the compact source of emission, if present. Because the surrounding cloud sometimes led to contamination, we primarily determine whether a source is detected upon visual inspection. Doing this, we find that 31 targets are detected in $^{12}$CO, 13 in $^{13}$CO, and 4 in C$^{18}$O. Gas is the main component of the disk mass and one way to determine the gas mass is the combination of $^{13}$CO and C$^{18}$O gas lines (e.g., \citealt{williams&best14}). In this sample only three objects have both $^{13}$CO and C$^{18}$O detections, thus we do not report gas masses for this small subset. The gas fluxes are presented in Table~\ref{tab: gas properties}.

\begin{figure*}[t]
    \centering
    \includegraphics[scale=0.56]{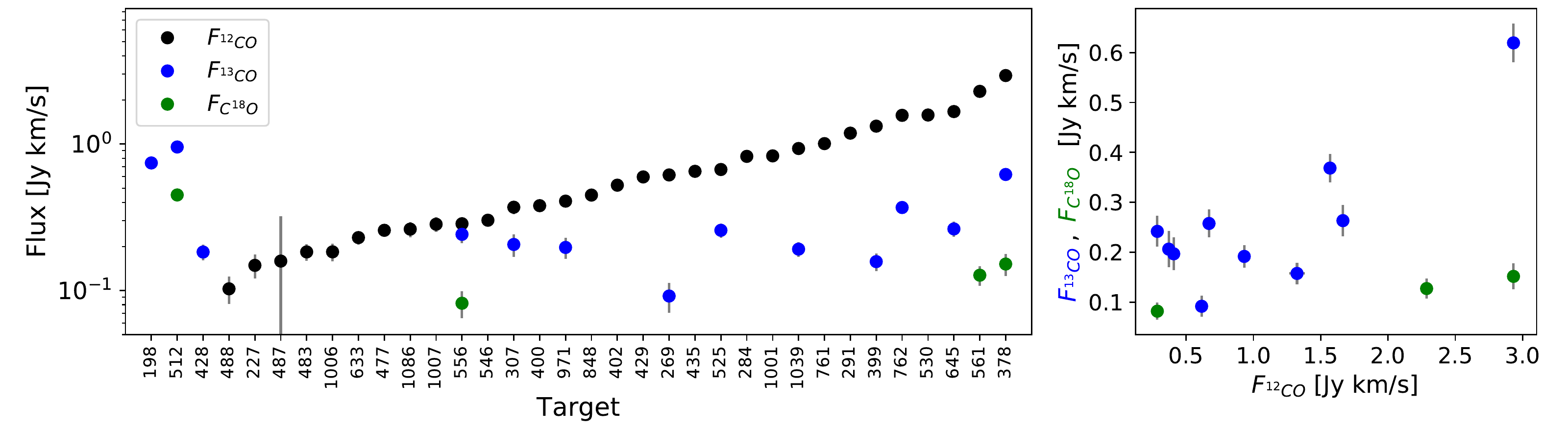}
    \caption{The CO gas fluxes for the targets where gas is detected. Left: The objects are ordered by $^{12}$CO flux (black). The $^{13}$CO and C$^{18}$O detections are in blue and green, respectively. [MGM2012] 198, 428, and 512 are not detected in $^{12}$CO, however they do show disk emission in $^{13}$CO and/or C$^{18}$O. Right: The $^{13}$CO (blue) and C$^{18}$O (green) fluxes as a function of $^{12}$CO flux. There is a strong correlation between the $^{12}$CO and $^{13}$CO flux (r=0.82) and a strong correlation between the $^{12}$CO and C$^{18}$O flux (r=0.99), although the latter has too few data points to consider this a reliable correlation.}
    \label{fig: gas flux}
    % from Analysis_gas.py
\end{figure*}

\begin{figure*}
    \centering
    \includegraphics[scale=0.72]{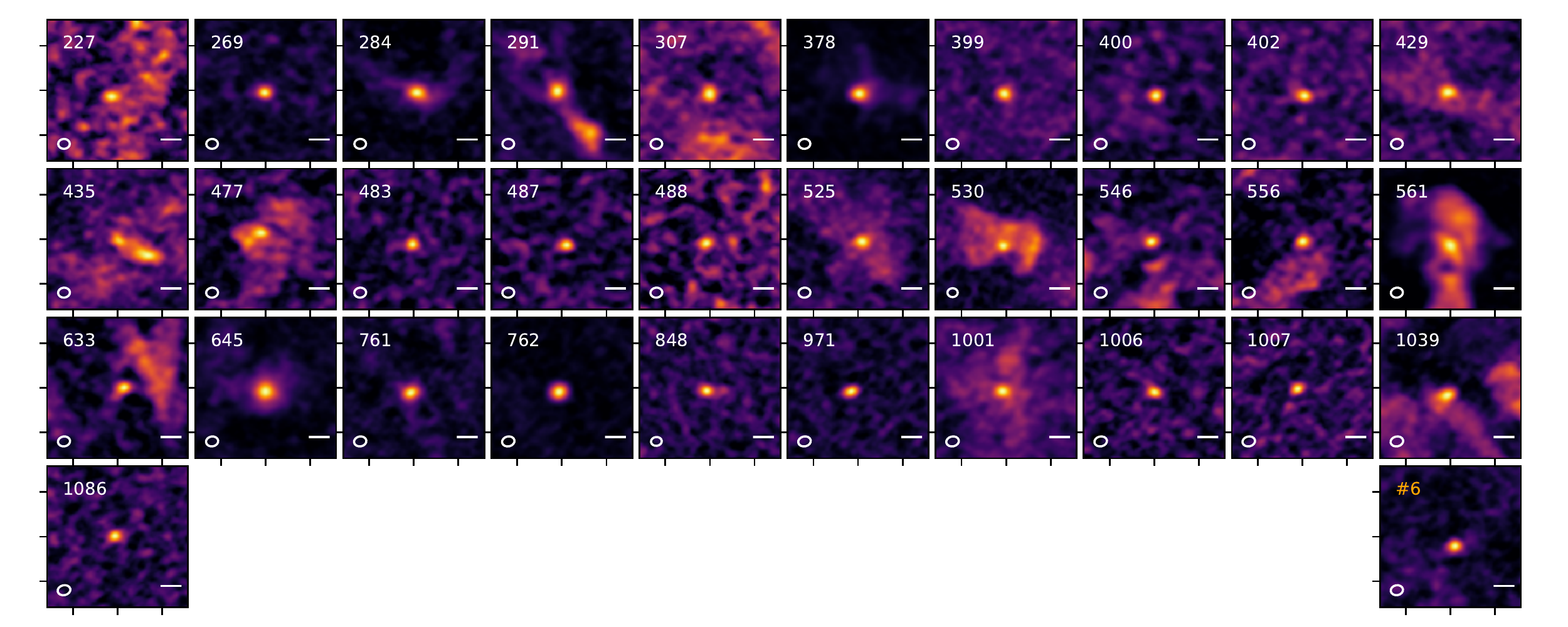}
    \caption{The $^{12}$CO zero-moment maps for the detected sources. Each image is 16''$\times$16''. A scalebar of 1000 AU is shown in the bottom right and the beam size is shown in the bottom left. The additional object that is detected in the $^{12}$CO zero-moment maps is shown in the bottom right with an orange label. The number corresponds to the numbers in Table~\ref{tab: extras table}.}
    \label{fig: 12COgas}
    % from FitsImages_gas.py
\end{figure*}

\begin{figure*}
    \centering
    \includegraphics[scale=0.72]{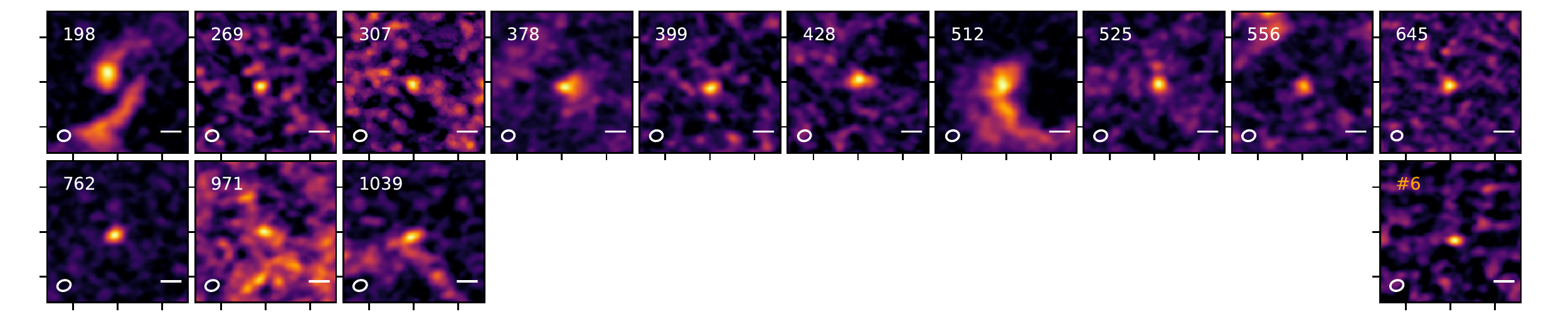}
    \caption{Same as Figure~\ref{fig: 12COgas} but for $^{13}$CO. }
    \label{fig: 13COgas}
    % from FitsImages_gas.py
\end{figure*}

\begin{figure}
    \centering
    \includegraphics[scale=0.72]{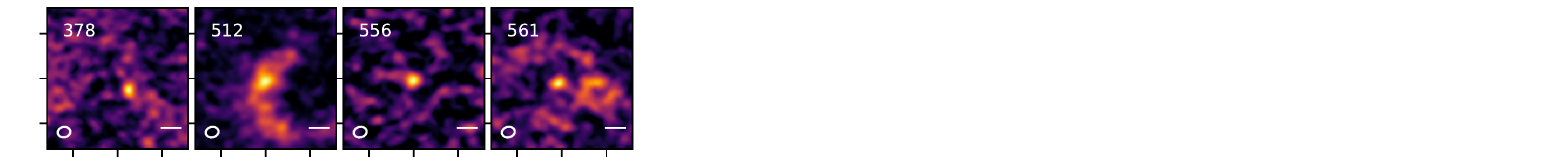}
    \caption{Same as Figure~\ref{fig: 12COgas} but for C$^{18}$O.}
    \label{fig: C18Ogas}
    % from FitsImages_gas.py
\end{figure}

In Figure~\ref{fig: gas flux} we show the gas fluxes for each detected target in order of increasing $^{12}$CO flux (left). In the right panel of Figure~\ref{fig: gas flux}, we show the $^{13}$CO and few C$^{18}$O fluxes as a function of the $^{12}$CO flux. We find a Pearson correlation coefficient of 0.82 between the $^{12}$CO and $^{13}$CO flux, indicating a strong correlation. Between the $^{12}$CO and C$^{18}$O flux, we find a coefficient of 0.99, indicating a strong correlation, however there are too few C$^{18}$O detections to determine if this is a real correlation. We find a moderate to strong correlation coefficient of 0.56 between the $^{12}$CO and continuum flux. There is no correlation between the $^{13}$CO and C$^{18}$O fluxes and the continuum flux. The zero-moment maps of the detected sources are shown in Figures~\ref{fig: 12COgas}, \ref{fig: 13COgas}, and \ref{fig: C18Ogas}. The gas shows extended emission for many objects, therefore we do not report gas radii. For some of these objects, the extended emission may be due to cloud contamination.

\subsection{[MGM2012] 512}\label{subsec: 512}
\begin{figure*}
    \centering
    \includegraphics[scale=0.6]{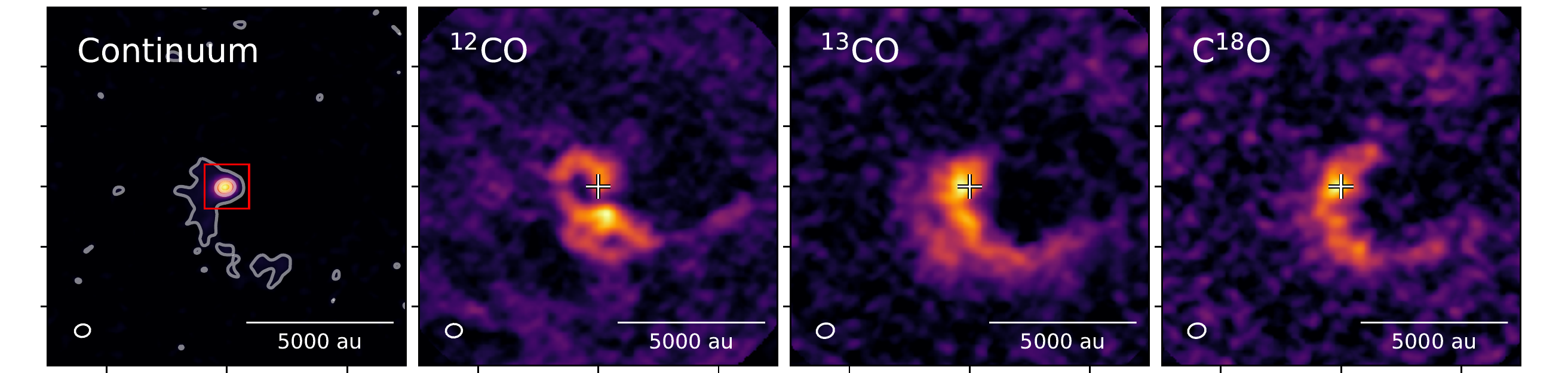}
    \caption{[MGM2012] 512 shows a large-scale structure extending to the south. We see this feature in the dust continuum, $^{12}$CO, $^{13}$CO, and C$^{18}$O (left to right). The beam is shown in the bottom left and a scale bar of 5000 AU is shown in the bottom right of each image given a distance of 410 pc to this object. The image size is 30''$\times$30''. In the continuum image, the red box is the region used to get the 1.33 mm continuum flux. In the gas zero-moment images the disk location is marked with a white plus sign.}
    \label{fig: 512}
    % from OddObjects.py
\end{figure*}

[MGM2012] 512 shows signs of large-scale structure in the continuum and $^{12}$CO, $^{13}$CO, and C$^{18}$O data. [MGM2012] 512, an M5.5$\pm$0.9 \citep{hsu12} object, shows a large arc extending from the disk to the south (Figure~\ref{fig: 512}). In $^{12}$CO we also see a lack of emission near the disk location, perhaps due to self-absorption. In Figure~\ref{fig: 512}, the red box in the continuum image is the region used to get the continuum flux reported in Table~\ref{tab: observations table} which does not include the extended emission. 

We determine continuum and $^{13}$CO and C$^{18}$O fluxes for the extended emission. Using the same box used to determine the disk flux and a custom region drawn by eye around the extended emission ($\sim$5" to the east, $\sim$9" to the south, and $\sim$6" to the west), the integrated continuum flux in those combined regions is 23.9$\pm$0.8 mJy. The background rms is $\sim$0.113 mJy beam$^{-1}$.  If we subtract the disk source (16.06$\pm$0.23 mJy), we get a flux of 7.8$\pm$0.8 mJy for the extended emission alone. In the  $^{13}$CO and C$^{18}$O zero-moment maps, we use similar apertures as for the continuum and find that the extended emission has a flux of 4.9$\pm$0.2 Jy km/s and 2.1$\pm$0.1 Jy km/s for the $^{13}$CO and C$^{18}$O, respectively. Using Equation~\ref{eq 1}, the continuum flux results in a dust mass for the extended emission of 65$\pm$32 \Mearth, $\sim$50\% of the disk mass. If this extended emission is envelope material, then the envelope-to-disk ratio is comparable to protostellar ratios and would put it in the Class I category found by \cite{jorgensen09}. Of the seven VANDAM protostars that show substructure, \cite{sheehan20} find that four have envelope-to-disk mass ratios that are lower than what we find here. Thus, if this is envelope emission, it is still quite massive relative to the disk. However, we note that Equation~\ref{eq 1} may not be appropriate for determining the dust mass within the arc. The disk mass may be underestimated because the disk may be optically thick, while the extended emission may have smaller grains, making the opacity used here too high, making the arc's mass also a lower limit. In the future, radiative transfer modeling of this object may be necessary to characterize the extended emission more fully, similar to \cite{sheehan20}. We discuss this large-scale emission and potential origins in Section~\ref{subsec: 512 disc}.

\section{Discussion}\label{sec: discussion}

\subsection{Comparing L1641 to the rest of Orion}\label{subsec: Orion disc}
Orion contains our nearest giant molecular cloud complex \citep{bally08} and is the benchmark for understanding star formation in giant molecular clouds due to its proximity and stellar and environmental diversity. Orion is also likely representative of other star-forming regions in the Milky Way, while closer star-forming regions lack the massive stars that many regions contain. It contains older regions as well as very young regions, dense populations, and clusters as well as low-density regions, and it has massive stars, like those in the Trapezium, that have a great impact on the nearby disks. However, Orion is also large enough that there are populations far from irradiating massive stars. There are several (sub-)millimeter surveys of different regions in Orion. \cite{eisner18} find disks with low masses and truncated radii in the ONC while \cite{vanTerwisga19} find that the OMC2 region, north of the ONC with lower stellar densities, has disk populations that are indistinguishable from low-mass systems like Taurus and Lupus. L1641 is estimated to have over 1600 Class II and Class III objects, comparable in size to the ONC but in a less dense region and with a lower radiation background \citep{hsu12,pillitteri13,megeath16}. We find that L1641 has a dust mass distribution that lies at higher masses than the young ONC \citep{eisner18} and older $\sigma$ Ori \citep{ansdell17}, similar to that of OMC1 \citep{eisner16}, and slightly lower masses than OMC2 \citep{vanTerwisga19}. This is further evidence that the dust mass distributions depend on the age of the region, evolutionary state, and exposure to external photoevaporation and stellar densities. However, our survey does not represent L1641 as a whole due to our bias towards infrared-bright disks that lie along the L1641 filament. Therefore, a comparison to other regions should be viewed with caution.

Our L1641 sources represent some of the brightest, youngest, and most massive Class II objects in L1641 that may bridge the gap between protostars and more evolved Class II systems. Our targets lie largely along the dense L1641 filament in the protostar-rich environments that were targeted with HOPS. The protostar/pre-main-sequence star ratio decreases from clustered to isolated environments \citep{megeath16} and our targets are largely in these clustered regions. Thus, it is likely that our sample is younger compared to a sample of disks that were either formed in isolation or have dispersed from higher density regions. The Orion VANDAM survey \citep{tobin20} targeted protostars in Orion with VLA and ALMA. The L1641 protostars can then be compared to the more evolved Class IIs, with no environmental factors complicating the comparison. 

The L1641 median dust mass (11.1$^{+32.9}_{-4.6}$ \Mearth) is lower than the median of the VANDAM protostellar disk sources in L1641 (85.1$^{+303.3}_{-13.5}$, 33.5$^{+154.2}_{-9.4}$, and 33.3$^{+138.6}_{-7.1}$ \Mearth\ for the Class 0, I, and Flat Spectrum sources, respectively, using $T_{dust}$=20 K and $\kappa_{\nu}$=3.45 cm$^{2}$ g$^{-1}$). Using a luminosity dependent dust mass ($T_{dust}\propto L_{bol}^{0.25}$) as in \citet{tobin20}, these median dust masses are 11.2$^{+45.8}_{-2.8}$, 7.4$^{+26.2}_{-2.6}$, and 7.5$^{+27.7}_{-1.8}$, for the Class 0, I, and Flat Spectrum sources, respectively, making the Class I and Flat Spectrum median dust masses lower than, but within the errors of, the Class II objects in this sample. This indicates that although the L1641 sample presented here may be biased towards massive disks, they still tend to have lower masses than their evolutionary predecessors, given a uniform temperature. This is also found for disks in Ophiuchus \citep{williams19}. However, for their evolutionary class, the L1641 Class II disks in our sample are quite massive with 27 out of 101 objects (27\%) having dust masses equal to or greater than the 30 \Mearth, the estimated amount needed to build the solar system \citep{weidenschilling77}. We note that the Class IIs in our sample and the protostars in the VANDAM survey were part of the HOPS observations and were originally identified by \cite{megeath12} based on spectral indices and photometry out to 24 \mic\ (see \citealt{megeath12} for full details), therefore disk mass was not a factor in the classification. Additionally, these masses may be underestimated given that the disk emission is optically thick (e.g., \citealt{ribas20, macias21}).

Several ALMA surveys of younger systems, Class 0/I and embedded systems, suggest that the dust mass reservoir decreases significantly as the systems evolve to the Class II stage. \cite{tychoniec20} compares a survey of Class 0 and Class I disks in Perseus to exoplanet solid masses and determines that a planet formation efficiency of 15\% is needed to form the observed planetary systems in the Class 0 phase and that efficiency increases to 30\% in the Class I phase. Following this trend, it may be too difficult for planet formation to begin in the Class II phase. \citet{grant18} found that L1641 showed signs of significant dust evolution based on the \textit{Herschel} fluxes in combination with irradiated accretion disk models \citep{d'alessio98,d'alessio06}. Despite dust evolution, and likely dust growth, the L1641 dust masses are still high.  In fact, the median dust mass of this sample, 11.1 \Mearth, is within the range for the Class 0 and I disks in Perseus (\citealt{tychoniec20}, see their Figure 6). Therefore, it may be feasible for the massive disks in L1641 to still have the mass reservoir needed to form giant planets, even at a lower efficiency.

\subsection{Transitional Disks in L1641}\label{subsec: TD disc}
Transitional disks were initially thought to be the intermediate stage between Class II disks and the less massive Class III disks. The ALMA surveys of Lupus by \cite{ansdell16} and \cite{vandermarel18} showed evidence that the transitional disk population is bright at millimeter wavelengths, indicating massive disks. \cite{vandermarel18} suggest that there may be a different evolutionary pathway for these massive transitional disks: massive disks form rings due to planets and/or condensation fronts and large cavities form from those rings. Less massive disks may go through a similar process, just on smaller scales, but without the mass necessary to form giant planets, or may alternatively have another evolutionary process. We find that the transitional disks in this sample have a median dust mass higher than the sample as a whole (16.1 \Mearth\ vs. 11.1 \Mearth), However, the samples are statistically similar. \cite{vandermarel18} emphasize the need to account for the dust cavity when converting from millimeter flux to dust mass, as not doing so can lead to overestimates. Therefore, the masses derived here for the transitional disks may be overestimated. \cite{vandermarel18} also find that four transitional disks in their sample would not be classified as such from their spectral energy distributions. The transitional disks in this sample were identified in \cite{grant18} using the spectral energy distributions, thus there are likely additional transitional disks in this sample. Surveys such as DSHARP \citep{andrews18} have shown how ubiquitous annular substructures are in disks and likely many of the disks in this region would show substructure at high angular resolution. The transitional disks referred to in this work have significant dips in their spectral energy distributions, which may indicate that these gaps or cavities are radially large to manifest in such a dip. Future observations of L1641 with higher angular resolution will therefore be important to characterize the gaps in these disks and to search for additional transitional disks with smaller gaps and/or cavities.

\subsection{The Disk Mass-Accretion Rate Relationship}\label{subsec: mdot-disk mass}
\begin{figure}
    \centering
    \includegraphics[scale=0.53]{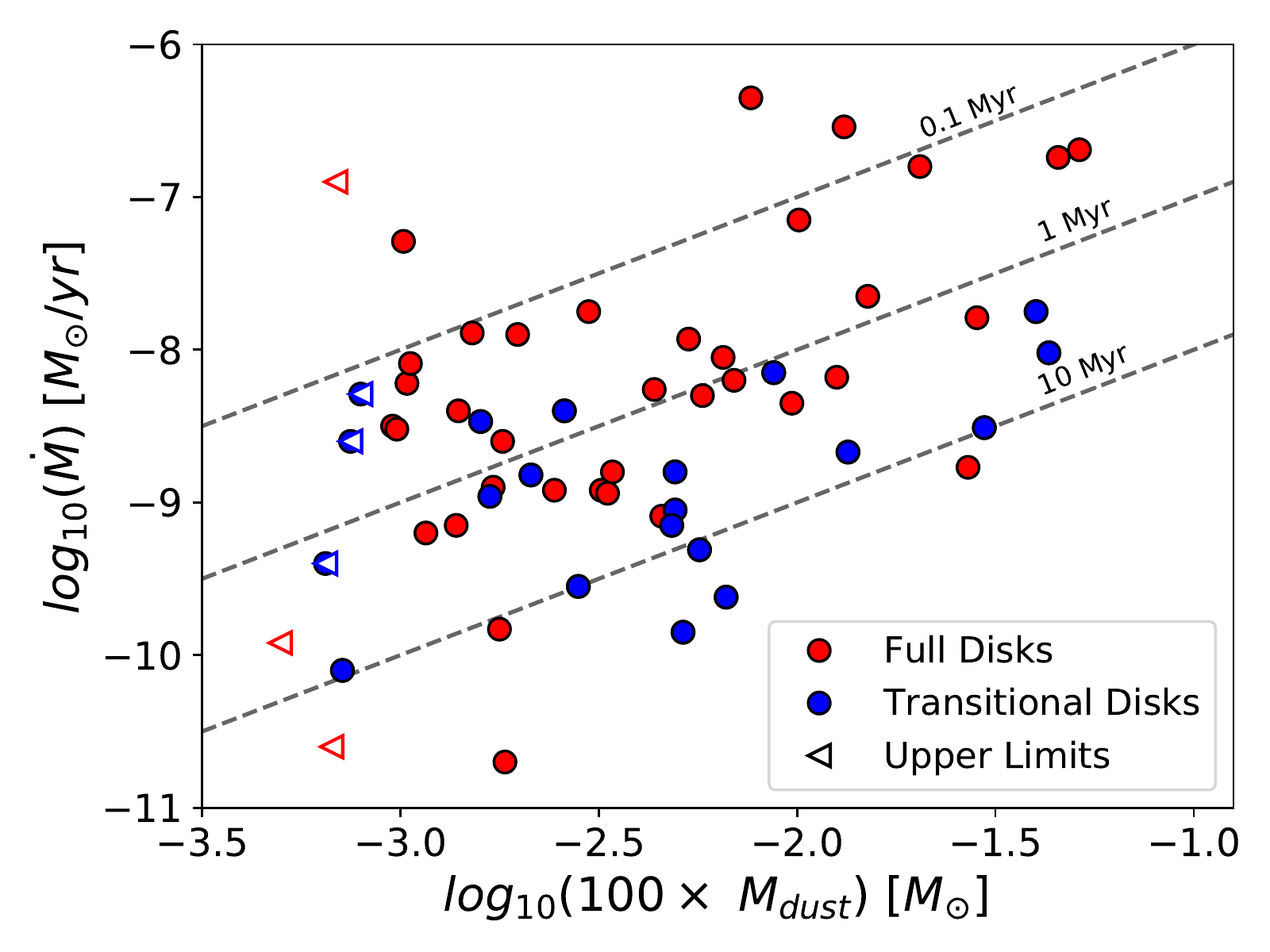}
    \caption{The mass accretion rate vs. disk mass (assuming a gas-to-dust mass ratio of 100). The full disks are shown in red and the transitional disks are shown in blue. Objects that are not detected in our continuum observations are shown as the open arrows. The labeled dashed lines show different timescales for viscous evolution ($M_{disk}/\dot{M}$).}
    \label{fig: Mdisk-Mdot}
    % from Analysis.py
\end{figure}

Recent ALMA surveys have provided dust masses for large samples of protoplanetary disks which have been combined with accretion rate surveys to constrain the disk mass-accretion rate relationship (e.g., \citealt{manara16,ansdell17,rosotti17,mulders17,lodato17,manara20}). For viscously evolving disks, the disk mass and accretion rate ($\dot{M}$) are related \citep{hartmann98}. This relationship can then be used to determine the viscous lifetime of the disk, $t_{disk}=M_{disk}/\dot{M}$ (e.g., \citealt{jones12,manara16,sellek20}). Figure~\ref{fig: Mdisk-Mdot} shows the disk mass (assuming a gas-to-dust mass ratio of 100) vs. accretion rate for our sample. The 58 accretion rates for our sample are obtained from \cite{fang09,fang13} and \cite{kim16}. The full sample has a correlation (in log-log space) of 0.44 and the transitional disks are remarkably similar with a correlation of 0.42. Therefore, while there is considerable scatter, there is a moderate correlation between $M_{disk}$ and $\dot{M}$ in our sample. The dashed lines correspond to the viscous lifetimes of the disks given different ratios of $M_{disk}/\dot{M}$. Overall, the viscous timescales for this sample are consistent with the age of the region ($\sim$1.5 Myr), however with a large scatter. The scatter that we find is not consistent with a single $\alpha$ from viscous evolution. There are several explanations for this scatter, including accretion variability, distance uncertainties (which would affect both $M_{disk}$ and $\dot{M}$, however different distances were used to obtain these two quantities), the effects of photoevaporation and planet formation on accretion, and potentially, the effects of infall from any remaining envelope material. Additionally, if the gas-to-dust ratio deviates from 100, the disk mass will be different, affecting the derived disk lifetime.

\subsection{Structure Around [MGM2012] 512}\label{subsec: 512 disc}
Here, we discuss potential origins of the large-scale structure that we observe coming off of the disk of [MGM2012] 512. These include emission from an envelope/accretion streamer or outflow cavity, or perturbations by an unseen companion. This arc is similar to the streamer that is feeding material from the local environment onto Per-emb-2 \citep{pineda20}, [BHB2007] 1 \citep{alves20}, SU Aur \citep{ginski21}, and, on a smaller scale, HL Tau \citep{yen19}. However, what is unique about the emission around [MGM2012] 512 is that it is detected in the continuum. This means that the dust grains are at least 100 \mic, larger than would be expected in the interstellar medium. If this emission is an accretion streamer, this could mean that the dust grains are able to grow as the material infalls onto the envelope and/or disk. 

This arc is qualitatively similar to the hydrodynamic modeling of \cite{dullemond19} and \cite{kuffmeier17,kuffmeier20} wherein low mass cloud fragments interact with a young stellar object. In some cases this cloud material may be accreted or may fly-by and interact only if the impact parameter is small enough. If this is the case, the accretion of this arc material may lead to an accretion outburst. It is also qualitatively similar to the dusty non-ideal magnetohydrodynamic model of \cite{lebreuilly20}. These authors find that dust grains settle efficiently, even in the pseudo-disk of inflowing streamers, agreeing with the presence of large grains that we observe in our ALMA observations.  

One potential source of the spiral formation could be due to an interaction with a nearby stellar system, however, no close companion to [MGM2012] 512 is visible in the 2MASS images. One object (2MASS 05401404-0731262) is $\sim$50 arcseconds to the north. If 2MASS 05401404-0731262 is at the same distance as [MGM2012] 512 (410 pc) then it would be 20,500 AU away, likely too far to be interacting. Additionally, there could be an unseen companion that is either perturbing from the outside or is being ejected from the system and disrupting the material as it moves outward from the system. 

\begin{figure}
    \centering
    \includegraphics[scale=0.5]{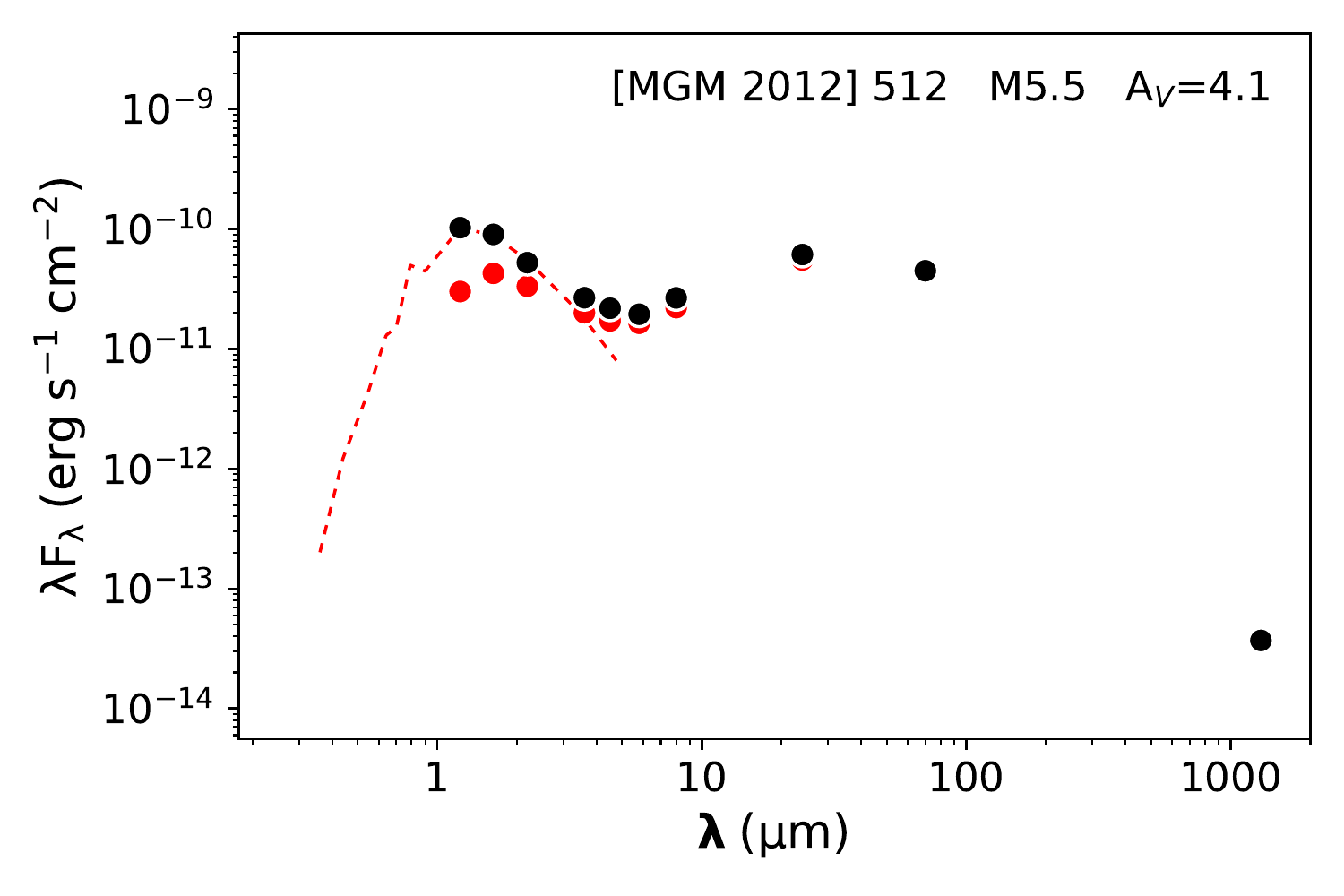}
    \caption{The spectral energy distribution of [MGM2012] 512. The red points are the observed photometry and the black points are the dereddened data given a visual extinction of 4.1 mag. The red-dashed line is the photosphere for an M5.5 star from \cite{k&h95}, scaled to the J-band. There is a large excess at mid- and far-infrared wavelengths, potentially due to envelope emission. }
    \label{fig: 512 sed}
    % From SEDS_individual_v2.py using the plots in SEDs/WithReddened/
\end{figure}

The spectral energy distribution of this target (Figure~\ref{fig: 512 sed}, with auxiliary data from the Two Micron All-Sky Survey, \textit{Spitzer} IRAC and MIPS, and \textit{Herschel} PACS, \citealt{skrutskie06, megeath12, grant18}) shows photospheric emission out to $\sim$3.6 \mic, after which it rises to have a large excess at mid- and far-IR wavelengths. This could be indicative of an envelope around this object, however we cannot rule out cloud contamination. In particular, this spectral energy distribution resembles the late Class I model in \cite{whitney03b} (Figure 3 in that work), indicating that this may be a protostar that is viewed pole-on, with the near-infrared data tracing the photosphere and the mid- and far-infrared tracing the combined disk and envelope emission. 
Although this is considered a young star with a disk (as opposed to a protostar) in \cite{megeath12}, \cite{carattiogaratti12} classifies this target as being a Class I object using the spectral index between 2.2 and 24 \mic. These authors classify objects as Class I if the spectral index is $\geq$0.3 and this object (ID 12 in their sample) has an index of 0.33. \cite{megeath12} apply the criteria of \cite{kryukova12}, and require a spectral index greater than -0.3 between 4.5 and 24 \mic\ and between 3.6 and 4.5 \mic.  Due to the more negative slope between 3.6 and 4.5 \mic\ for this source, \cite{megeath12} classified this source as a pre-main-sequence star with a disk. One issue is that the 2.2-24 \mic\ slope used by \cite{carattiogaratti12} is sensitive to the effects of foreground extinction. Nevertheless, both criteria show that this source is very close to the dividing lines used to distinguish between protostars and pre-main-sequence stars with disks. Special cases, such as protostars with face-on inclinations, or transitional phases between protostars and pre-main-sequence stars, may give discrepant results using these criteria. This object clearly shows interesting behavior that is outside of the scope of this work. We leave additional analysis to follow-up efforts.

\clearpage

\section{Summary and Conclusions}\label{sec: summary and conclusions}
We present an ALMA survey of protoplanetary disks in Lynds 1641. The 101 disks studied here represent some of the far-infrared-brightest Class II objects in the region, making it a valuable sample to compare to other star-forming regions and to the protostellar sample that co-exists in L1641. These observations are taken in Band 6 which covers the continuum at 1.33 mm and the $^{12}$CO, $^{13}$CO, and C$^{18}$O J=2-1 lines. Our sample is biased by the requirement that these objects were observed and detected at 70 \mic\ as part of the \textit{Herschel} Orion Protostar Survey. This survey focused on detecting far-infrared-bright protostars and our Class II sample was serendipitously observed. This indicates that our sample is biased towards the brightest disks that lie along the dense L1641 filament. Our results include the following: 

\begin{enumerate}
    \item We detect 89/101 (88\%) in continuum at the 4$\sigma$ level. We use the continuum fluxes to get dust masses or upper limits for all 101 disks in our sample. In the CO gas data, we detect 31 in $^{12}$CO, 13 in $^{13}$CO, and 4 in C$^{18}$O. 
    
    \item A fraction (27\%) of our sample has a dust mass equal to or greater than the minimum mass solar nebula dust mass value of $\sim$30 \Mearth. We find a median dust mass of 11.1$^{+32.9}_{-4.6}$ \Mearth\ for the L1641 Class II disks.
    
    \item The Orion VANDAM survey \citep{tobin20} allows for comparison between our Class II sample to Class 0/I/Flat Spectrum protostars in the same region. We find that the median dust mass of our sample is lower than those found for the protostars. Thus, L1641 Class II objects follow the trend of having lower dust masses than their younger counterparts, even with our bias to far-infrared-bright disks. 
    
    \item We observe 23 transitional disks (identified by their spectral energy distributions in \citealt{grant18}), 20 of which are detected in the continuum, which have a median dust mass of 16.1 \Mearth. Of the 36 disks detected in the gas, 10 are transitional disks.
    
    \item The disk mass-accretion rate relationship in our sample is largely consistent with viscous accretion timescales, however, there is a large scatter in the relationship. This may be due to accretion variability, systematics, and/or photoevaporation and disk evolution effects. 
    
    \item One object, [MGM2012] 512, shows large-scale structure in the dust continuum and all three gas lines. If this material is remnant envelope material, then the envelope-to-disk dust mass ratio is $\sim$50\%, on par with protostellar systems. This could be an accretion streamer connecting the local environment and the disk. The fact that this structure is detected in the continuum indicates relatively massive dust grains.
    
\end{enumerate}

This survey introduces a sample of protoplanetary disks at a crucial stage of evolution. A complete ALMA survey of L1641 in the future will help determine how this subset fits into the region as a whole. Paired with the complete sample of protostellar disks in L1641 in the Orion VANDAM survey, this region has great potential to test disk evolution theories. 

\acknowledgments
We thank the referee for a careful reading of the paper and for comments that improved the manuscript. We also thank Marina Kounkel, Sierk van Terwisga, Dominique Segura-Cox, Michael K{\"u}ffmeier, and Ugo Lebreuilly for insightful discussions that helped the manuscript. We acknowledge support from the NRAO Student Observing Support program through award SOSPA7-015. This paper makes use of the following ALMA data: ADS/JAO.ALMA\#2019.1.00951.S. ALMA is a partnership of ESO (representing its member states), NSF (USA) and NINS (Japan), together with NRC (Canada), MOST and ASIAA (Taiwan), and KASI (Republic of Korea), in cooperation with the Republic of Chile. The Joint ALMA Observatory is operated by ESO, AUI/NRAO and NAOJ. The National Radio Astronomy Observatory is a facility of the National Science Foundation operated under cooperative agreement by Associated Universities, Inc. 

\bibliographystyle{aasjournal}
\bibliography{biblio}

\end{document}